\documentclass[a4paper,11pt]{article}
\pdfoutput=1 % 

\usepackage{jinstpub} % 

\usepackage{lineno}
\usepackage{graphicx}% Include figure files
\usepackage{dcolumn}% Align table columns on decimal point
\usepackage{bm}% bold math
\usepackage{adjustbox}

\graphicspath{{ps/}}
\usepackage{subfigure}
\usepackage{epstopdf}
\usepackage{amsmath}
\usepackage{lineno,hyperref}
\usepackage{makecell}
\usepackage{multirow}
\usepackage{rotating}
\usepackage{enumitem}
\usepackage{url}
\usepackage{color,soul}
\modulolinenumbers[5]

\title{High intensity proton beam impact at 440 GeV/c on Mo and Cu coated CfC/graphite and SiC/SiC absorbers for beam intercepting devices}

\author[a,b]{J.~Maestre,}
\author[a]{C.~Bahamonde,}
\author[a]{I.~Lamas~Garcia,}
\author[a]{K.~Kershaw,}
\author[a]{N.~Biancacci,}
\author[a]{J.~Busom,}
\author[a,c]{M.~I.~Frankl,}
\author[a]{A.~Lechner,}
\author[a]{A.~Kurtulus,}
\author[d,e]{S.~Makimura,}
\author[f]{N.~Nakazato,}
\author[a]{A.~T.~P{\'e}rez,}
\author[a]{A.~Perillo-Marcone,}
\author[a]{B.~Salvant,}
\author[a]{R.~Seidenbinder,}
\author[a]{L.~Teofili}
\author[a,1]{and M.~Calviani\note{Corresponding author.}}

\affiliation[a]{European Organization for Nuclear Research \\ CERN, Esplanade des Particules 1, 1211 Genève 23, Switzerland}
\affiliation[b]{Vicerrectorado de Investigación y Transferencia \\ Universidad de Granada, Gran Vía de Colón 48, Granada, Spain}
\affiliation[c]{Paul Scherrer Institute, Forschungsstrasse 111, 5232 Villigen PSI, Switzerland}
\affiliation[d]{Muon Science Section, Materials and Life Science Division \\ J-PARC
Center, Tokai, Ibaraki, 319-1106, Japan}
\affiliation[e]{Muon Science Laboratory, Institute of Materials Structure Science \\
High Energy Accelerator Research Organization, Tokai, Ibaraki, 319-1106,
Japan}
\affiliation[f]{College of Information and Systems, Muroran Institute of Technology \\
27-1 Mizumoto-cho, Muroran, Hokkaido 050-8585, Japan}

\emailAdd{marco.calviani@cern.ch}

\date{\today}% It is always \today, today,
\abstract{
Beam Intercepting Devices (BIDs) are essential protection elements for the operation of the Large Hadron Collider (LHC) complex. The LHC internal beam dump (LHC Target Dump Injection or LHC TDI) is the main protection BID of the LHC injection system; its main function is to protect LHC equipment in the event of a malfunction of the injection kicker magnets during beam transfer from the SPS to the LHC. Several issues with the TDI were encountered during LHC operation, most of them due to outgassing from its core components induced by electron cloud effects, which led to limitations of the injector intensity and hence had an impact on LHC availability. The absorbing cores of the TDIs, and of beam intercepting devices in general, need to deal with high thermo-mechanical loads induced by the high intensity particle beams. In addition, devices such as the TDI - where the absorbing materials are installed close to the beam, are important contributors to the accelerator impedance budget. To reduce impedance, the absorbing materials that make up the core must be typically coated with high electrical conductivity metals. Beam impact testing of the coated absorbers is a crucial element of development work to ensure their correct operation. 

In the work covered by this paper, the behaviour of several metal-coated absorber materials was investigated when exposed to high intensity and high energy proton beams in the HiRadMat facility at CERN. Different coating configurations based on copper and molybdenum, and absorbing materials such as isostatic graphite, Carbon Fibre Composite (CfC) and Silicon Carbide reinforced with Silicon Carbide fibres (SiC-SiC), were tested in the facility to assess the TDI's performance and to extract information for other BIDs using these materials. In addition to beam impact tests and an extensive Post Irradiation Examination (PIE) campaign to assess the performance of the coatings and the structural integrity of the substrates, extensive numerical simulations were carried out. 
}


\arxivnumber{XXXX.YYYYY} % only if you have one

\begin{document}
\maketitle

\section{Introduction}
\subsection{Beam intercepting devices for the LHC}
Beam intercepting devices (BIDs) play a vital role in the Large Hadron Collider's (LHC) safe operation. Their functions include cleaning halo particles and physics debris, controlling the beam size, protecting equipment from damaging beam impacts and controlling beam losses of the system, hence radio-activation, around the complex. 

Proton beams injected into the LHC arrive from the Super Proton Synchrotron (SPS) accelerator via two transfer lines (TI2 and TI8). The beams are injected by means of septa and kicker systems into the LHC. Machine protection systems are installed in these areas to ensure protection of LHC elements in the event of mis-steered beams~\cite{Bruning:782076}. The LHC Internal Beam Dumps (LHC Target Dump Internal or TDI) are the main beam intercepting protection elements for the injection of the LHC~\cite{mete2014,Bruning:782076}. There are two TDIs in the LHC, one at each injection point (downstream of transfer lines TI2 and TI8).

There are different kinds of LHC proton beams depending on the filling scheme, but all of them are based on a specific bunch structure~\cite{papaphilippou2014}. A series of beams of up to $288$ bunches from the SPS at $450$ GeV of energy are injected into the LHC. If their trajectory deviates outside acceptable limits, the injected beam is intercepted by the TDI. The bunch intensity has been increased over the years of LHC operation, reaching $1.2\cdot 10^{11}$ protons per bunch (ppb) at the end of Run 2 (2015-2018). 

Malfunctions of the injector kickers could occur during injection~\cite{goddard2017}. This malfunction may cause off-nominal beam orbits and hence lead to a risk of interaction between the proton beam and sensitive LHC equipment, such as the cryogenic magnets close to the experimental insertions. The TDIs, placed at a phase advance of $75-95^{\circ}$ from the injection kickers ($70$ m downstream), intercept such off-trajectory beams. 

\begin{figure}[htpb]
\centering
\includegraphics[width=0.5\linewidth]{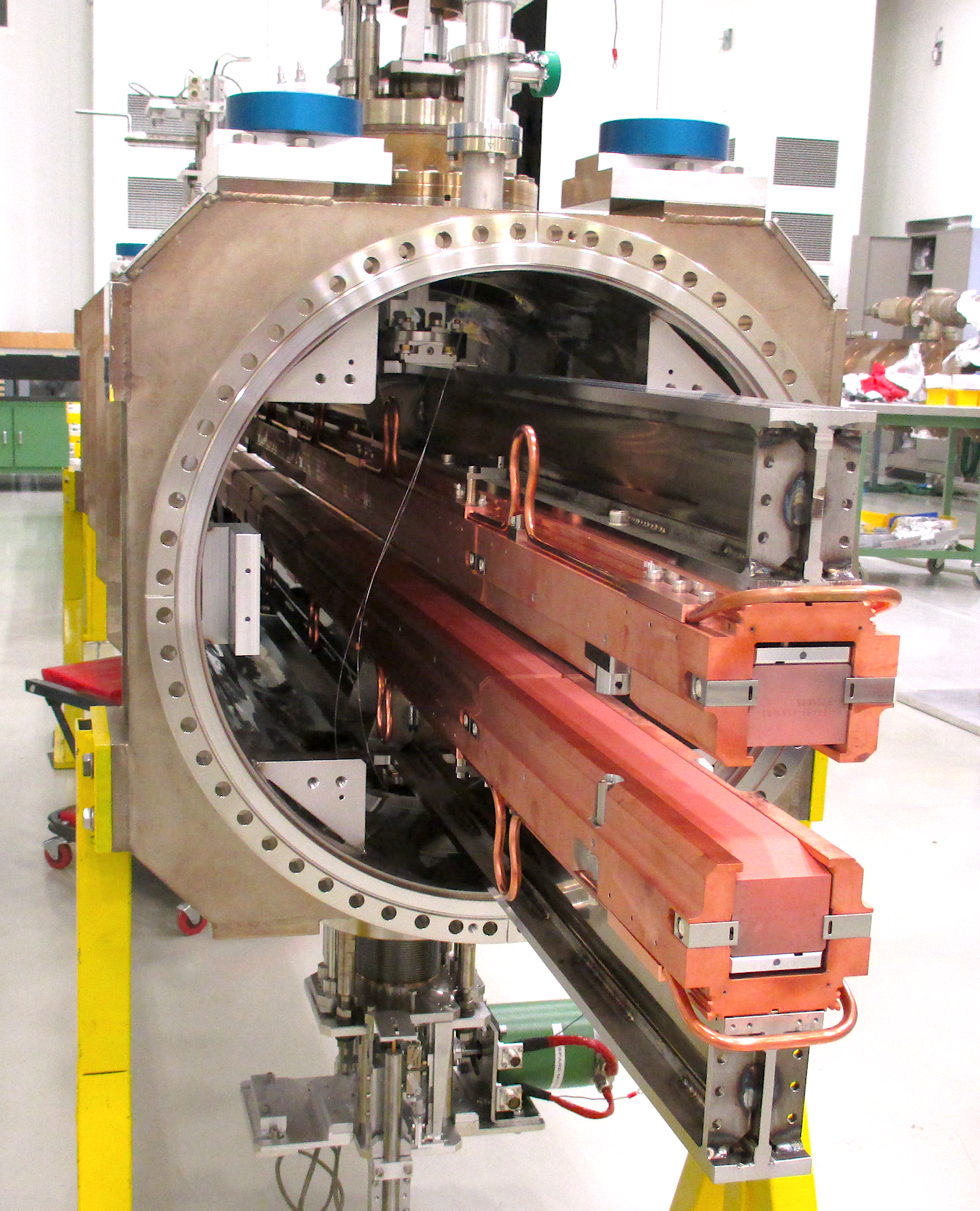}
\caption{The image shows the upstream end of the two assembled top and bottom  jaws of a TDI, during preparation in a clean-room prior to installation in the LHC. The Cu coating on the isostatic graphite and the supporting module is clearly visible.}
\label{fig:TDI}
\end{figure}

Each LHC TDI consists of a pair of $4.185$~m long, movable "jaws" inside a vacuum tank (see Fig.~\ref{fig:TDI}). Each jaw accommodates several absorbing blocks made of different materials, originally titanium-coated hexagonal boron nitride (hBN), aluminium (Al) and copper-beryllium (CuBe), in order to absorb the beam kinetic energy. During injection, the particle beam passes between the jaws, which are separated by a gap of $7.6$ mm ($6.8\sigma$, where $\sigma$ is the transverse beam size in the vertical plane), providing protection in the event of failure of the kickers whilst providing some operational margin in case of orbit variations~\cite{bracco2012}. At the end of the injection, the jaws are retracted to their parking positions (gap = 110 mm), allowing the ramp up of the LHC beam energy.

\subsection{TDI operation and design considerations}

Several injection failures occurred during the LHC Run~1 operation period (2009-2013), involving direct beam impact onto TDI jaws and in some cases grazing along the coating surface (one of the most harmful failures)~\cite{lechner2013,lechner2015}. These incidents unavoidably led to quenching of downstream magnets due to the secondary particle shower, but no equipment damage occurred as by design.  
Although the TDIs performed as expected, a series of anomalies were detected during this run~\cite{lechner2015,bracco2011}. In brief, these consisted of: excessive deformation of some metallic components resulting in a drift of jaw position during injection due to thermal beam-induced loads, deformations of the beam screens, and some vacuum issues, which increased the beam-induced background in the nearby ALICE experiment. Severe coating degradation was also observed, with potential consequences on impedance. Although many of these issues were resolved after the Long Shut Down~1 (LS1) upgrade, new issues arose during Run~2 (2015-2018) as the injection intensity was increased. Large vacuum spikes were observed during beam injections, resulting in premature beam aborts. In addition, laboratory tests revealed unexpected failures of the hBN blocks when heated up to temperatures much lower than the service temperature limit specified by the manufacturer. In order to reduce the risk of damage, the intensity of the injected beams was temporarily limited~\cite{lechner2015} until the hBN blocks were replaced by copper-coated (layer thickness = $2$~$\mu$m) isostatic graphite during the Year-End Technical Stop (YETS) of 2015-2016.

The absorber core of the TDI (see Fig.~\ref{fig:TDI}), as for other beam intercepting devices in general, is exposed to high thermo-mechanical loads induced by particle beams~\cite{simos2006,nuiry2019}. Two main criteria need to be considered for the design of absorbing materials: the thermal shock resistance~\cite{simos2016,Hurh2013} and, for devices close to the beam, the resistive wall impedance~\cite{doi:10.1142/8543,Day2013,Teofili_2018}. During LHC beam operation, the electromagnetic fields generated by the proton beam are perturbed when the passage in-between the jaws. This electromagnetic coupling, known as impedance, has important implications on the beam dynamics because it may generate beam instabilities and in addition it may induce additional thermal loads in the surrounding components. In order to reduce the impedance, the TDI absorbers are coated with high electrical conductivity materials (typically copper)~\cite{mereghetti2020,Biancacci2017}. In the event of direct beam impact into the absorber, a sudden energy deposition and increase of temperature is generated as a consequence of the beam-matter interaction: this leads to transmission stress waves in the material due to rapid thermal expansion. Absorber materials are therefore selected to withstand high thermal shock loads. In practice, however, the coating is usually the most vulnerable element as it could be damaged in the event of beam impact, compromising the impedance performance of the equipment and potentially creating other challenges.     

\subsection{Beam intercepting device material testing}
Testing of beam absorbing materials as part of the BID design and development process is crucial to ensure the correct operation of BIDs. With the aim of validating the TDI's integrity throughout its service lifetime and to address the general uncertainties about the behaviour of coated-absorbing blocks when exposed to high energy beams, an experimental set-up was designed and installed for testing in CERN's HiRadMat facility~\cite{harden2019,Harden:IPAC2019-THPRB085,Efthymiopoulos:1403043}, called hereinafter HRMT-35. This test was carried out in 2018. The main goal was to test and to validate the performance of coated graphitic blocks under direct beam impact conditions. Two different coatings were considered in the experiment, copper (similar configuration to the TDI jaws) and molybdenum. Although molybdenum has a lower electrical conductivity than copper, it was thought that it may provide better thermal performance due to its higher melting point. 

In addition to isostatic graphite, two other promising absorbing materials based on Carbon Fibre Composite (CfC) and Silicon-Carbide fibre reinforced Silicon-Carbide (SiC-SiC) were also investigated. The excellent thermo-mechanical properties of the carbon fibres enhance the performance of the CfC material, in particular in terms of strength, fracture toughness and fatigue resistance with respect to graphite. Currently, CfCs are being used in some BIDs subjected to extremely high thermo-mechanical conditions, such as LHC and SPS-to-LHC transfer line collimators~\cite{bertarelli2005,nuiry2019}. SiC-SiC is also thought to be a promising potential alternative as an absorbing or particle producing target material. In recent developments of this material for nuclear physics applications, SiC-SiC targets are being tested for muon/pion production~\cite{makimura2020}. SiC-SiC exhibits high thermal shock resistance and fracture toughness, and, unlike carbon-based materials, has good oxidation resistance. This latter quality is ideal in the event of vacuum problems and subsequent direct contact with oxygen at high temperatures, as it ensures its chemical stability. Nevertheless, the thermal shock response when exposed to high intensity proton beams needed to be assessed.

To the best knowledge of the authors, the HRMT-35 experiment represents the first tests of coated absorbing materials exposed to high intensity proton beams and it offers the opportunity to gain further knowledge of advanced absorbing materials with potential applications in beam intercepting devices. The present work describes the design and set-up of the HRMT-35 experiment (Sec.~\ref{Design_Experiment}), the expected thermo-mechanical response of the different tested materials via numerical analysis (Sec.~\ref{Num_Analysis}), the execution of the tests (Sec.~\ref{Exp_execuation}) and the experimental findings including a comprehensive Post-Irradiation-Examination (PIE) campaign and subsequent discussion based on the numerical results (Sec.~\ref{Results}). The main conclusions of this work are finally summarized in Sec.~\ref{Conclusion}.

\section{Experiment design and set-up}
\label{Design_Experiment}
\subsection{Experiment and target summary}
As part of the preparations for experiments in the HiRadMat facility, materials to be tested have to be carefully prepared and integrated in a handling module which is then installed in the beam line using the facility's crane. A key requirement for this work was that the experiment was executed in primary vacuum conditions ($10^{-3}$~$mbar$) to prevent oxidation of the carbon blocks at high temperature. For this reason all the material samples were installed inside a vacuum tank and the different material samples were installed on motorised supports inside the tank so that they could be individually remotely aligned with the proton beam to mimic operational conditions.

Four targets were irradiated during the HRMT-35 experiment. Each target consists of long parallepiped blocks made of different materials (substrate) and coatings. The four targets were:

 \begin{itemize}
 \item Target 1: Copper-coated isostatic graphite. 
 \item Target 2: Molybdenum-coated isostatic graphite.
 \item Target 3: Molybdenum-coated CfC.
 \item Target 4: SiC-SiC composite.
\end{itemize}\par

\begin{figure}[htbp]
\begin{center}
\includegraphics[width=0.6\linewidth]{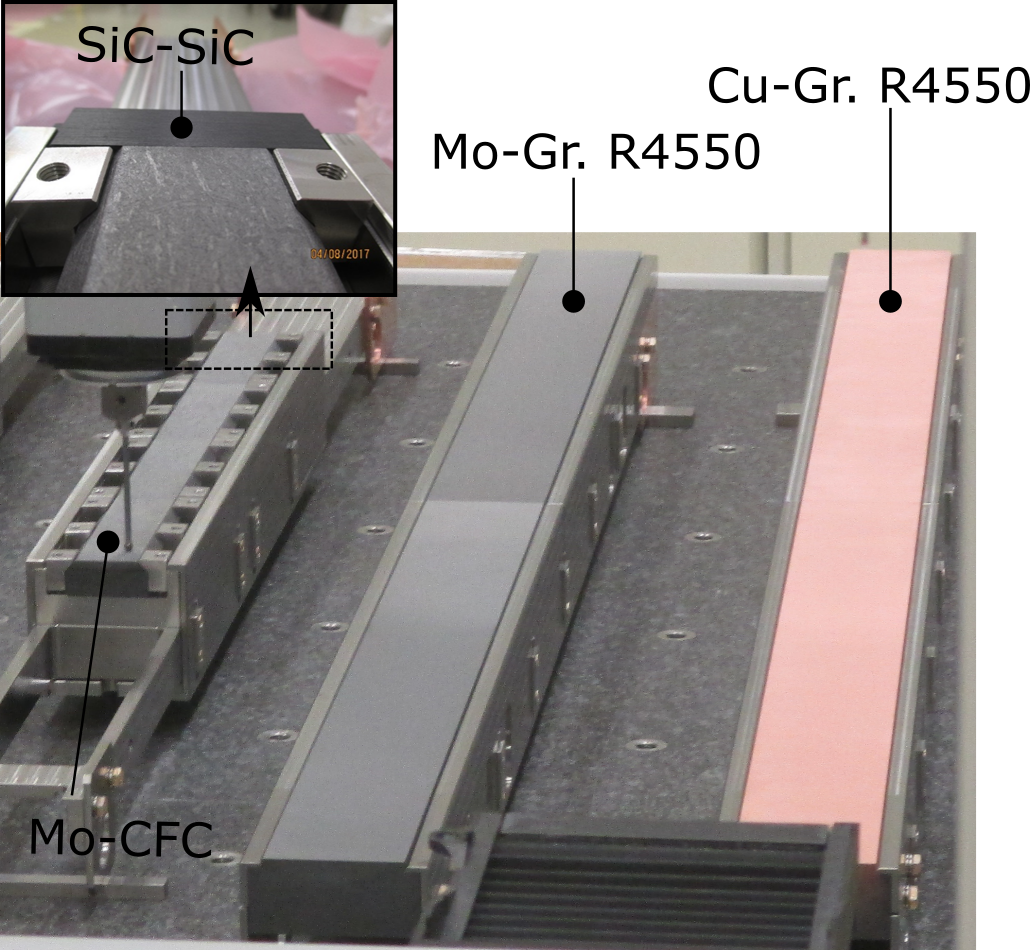}
\end{center}
 \caption{Photo of the targets prior to their installation in the HRMT-35 experimental tank. The absorber blocks are shown clamped in the jaws that guarantee their straightness to the required tolerances. It is worth noticing that the SiC-SiC sample was installed downstream the Mo-coated CfC absorber.}\label{fig1}
\end{figure}

\begin{figure}[htbp]
\begin{center}
\includegraphics[width=0.6\linewidth]{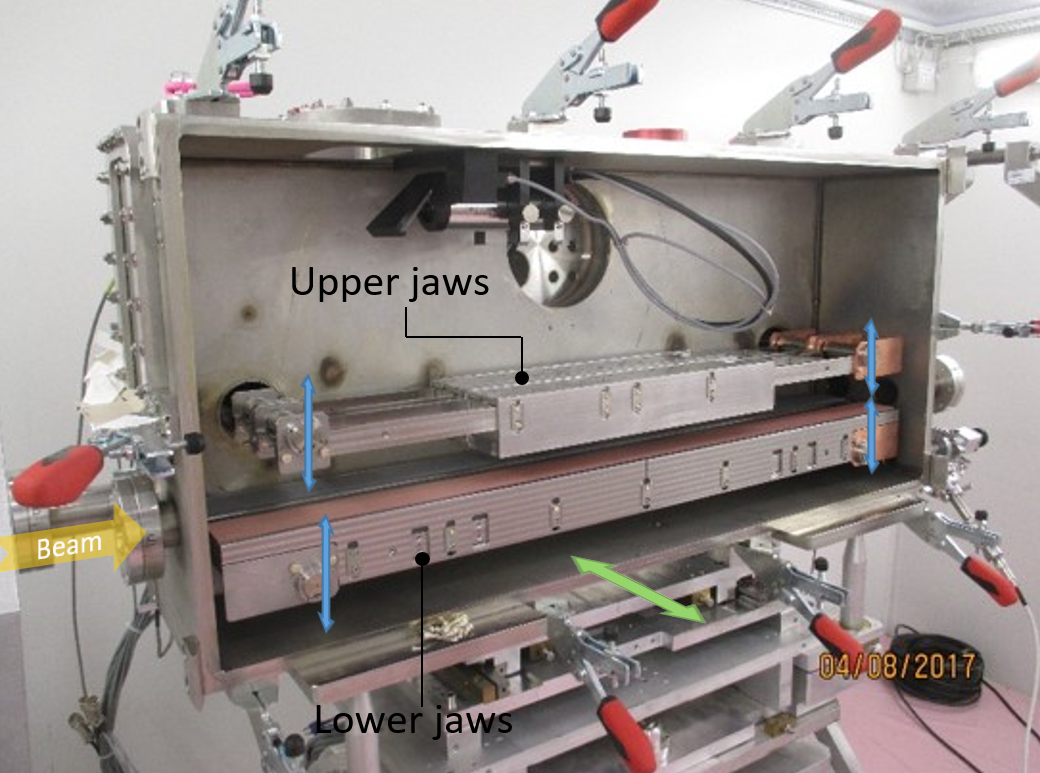}
\end{center}
 \caption{Open tank showing the blocks to be irradiated installed in the upper and lower jaws. The arrows represent the motorized movements of the system: vertical movements of each extremity of the jaw (blue arrows) with a range of $\pm30$~mm; and horizontal movement of the whole tank (green arrow) with a range of $\pm60$~mm.}\label{fig2}
\end{figure}

As shown in Fig.~\ref{fig1}, the targets were mechanically clamped in three supports, referred to as jaws. The jaws were assembled inside a vacuum-tight tank: two long jaws at the bottom of the tank and two smaller jaws above them (see Fig.~\ref{fig2}). %Note that the tank includes an extra jaw on the upper side used for another purpose out of the scope of present experiment. 
A summary of the targets and dimensions can be found in Table \ref{table1}. 

\begin{table}[htbp]
\centering
%\begin{adjustbox}{width=\columnwidth,center}
\begin{tabular}{c|cccc}
\multicolumn{1}{c|}{\begin{tabular}[c]{@{}c@{}}Target \\ position\end{tabular}} & \begin{tabular}[c]{@{}c@{}}Blocks\\ {[}quantity{]}\end{tabular} & \multicolumn{1}{c}{\begin{tabular}[c]{@{}c@{}}Dimension\\  {[}mm{]}\end{tabular}} & Substrate & \multicolumn{1}{c}{Coating} \\ \hline
\begin{tabular}[c]{@{}l@{}}Bottom  right\end{tabular} & 2 & 37x80x600 & \begin{tabular}[c]{@{}c@{}}SGL Graphite\\ grade R4550\end{tabular} & Copper \\
\begin{tabular}[c]{@{}l@{}}Bottom left\end{tabular} & 2 & 37x80x600 & \begin{tabular}[c]{@{}c@{}}SGL Graphite\\ grade R4550\end{tabular} & Molybdenum \\
\begin{tabular}[c]{@{}l@{}}Top right\end{tabular} & 4 & 20x45x125 & \begin{tabular}[c]{@{}c@{}}Tatsuno\\ FS 140\end{tabular} & Molybdenum \\
\begin{tabular}[c]{@{}l@{}}Top right\end{tabular} & 1 & 26x40x84 & SiC-SiC & \multicolumn{1}{c}{-}
\end{tabular}
%\end{adjustbox}
\caption{Target positions in the test tank and technical description and dimensions of the absorber blocks, as shown also Fig.~\ref{fig1} and \ref{fig2}.}\label{table1}
\end{table}

\subsection{Target materials and preparation}
Targets with the same substrate, isostatic poly-crystalline graphite at a density of $1.83$~g/cm$^3$, produced by SGL (grade R4550), were installed in both long bottom jaws, only differing in the coating material. One jaw was coated with copper, as installed in the TDI, and the other-one was coated with with molybdenum, a promising material for impedance reduction with a higher melting point than the copper - as installed in the HL-LHC collimators~\cite{coatings10040361}.

The upper jaw was fitted with four blocks of Carbon-Fibre-Composite (CfC) produced by Tatsuno Co. (grade FS~140) coated with molybdenum. This substrate material is made up of two-dimensional layers, piled-up through the thinnest dimension of the block (third direction) and with a random disposition of carbon fibres in the plane. The fibres are bonded by a graphitic matrix, highly graphitized at $2800$~$^\circ C$, leading to a final density of $1.85$~ g/cm$^3$. The result is a 2D-orthotropic material, with a virtually isotropic behaviour in the fibre plane~\cite{accettura2019,bianchi2017}. 

Also in the upper jaw, a SiC/SiC block, produced by the Organization of Advanced Sustainability Initiative for Energy System/Materials (OASIS) in Muroran institute of Technology~\cite{kohyama2011}, was placed downstream of the CfC blocks. The SiC/SiC composite was produced by the nano-infiltration and transient eutectic-phase (NITE) process. The NITE process is an applied liquid phase sintering method, which is appropriate to form the crystallized and dense SiC matrix in SiC/SiC composites. In comparison with  carbon-based materials, SiC-SiC has a significantly higher density ($2.83$~g/cm$^3$) and has a much greater oxidation resistance~\cite{PARK2018}. The tested SiC-SiC composite is built up in layers of $150-200$~$\mu$m thick made of alternating unidirectionally oriented fibres in a  $0/90$ degree configuration~\cite{makimura2020}. The block was oriented with layers perpendicular to the beam direction.

A Direct Current Magnetron Sputtering (DCMS) technique was used for the deposition of the thin coating on the different substrates (except the SiC-SiC block), with a nominal thickness of $2.5$~$\mu$m. Before applying the coating the substrates were cleaned in an ultrasonic bath and subsequently fired in order to remove impurities and hence to ensure a good adherence. In addition, the substrates were baked-out at $950$~$^{\circ}C$ under vacuum for a minimum of $2$ hours with the aim of reducing out-gassing into the primary vacuum during the experiment~\cite{coatings10040361}.

\subsection{Experimental tank and target supports}
As shown in Fig.~\ref{fig2}, the experiment was equipped with a motorization system, which provides three Degrees of Freedom (DoF) per jaw. This system permits independent movement of the extremities of each jaw in the vertical axis by means of accurate motor-resolvers (with a resolution of 5~$\mu$m/step). In addition, the tank can be moved horizontally perpendicular to the beam. The position of the jaws (with respect to a common reference) is given by the resolver and checked by Linear Variable Differential Transformer sensors (LVDT) with an accuracy of roughly $10$~$\mu$m. Using this system, each jaw can be accurately aligned with respect to the beam in order to be irradiated as required for the experiment (the alignment procedure will be explained in Sec.~\ref{Exp_execuation}).

\subsection{Beam impact parameters}
Similar energy densities to the ones that the TDIs face in operation can be provided at the HiRadMat facility~\cite{harden2019,Harden:IPAC2019-THPRB085,Efthymiopoulos:1403043}, as it is fed with high-intensity, LHC-like proton beams extracted from the SPS. Table~\ref{table2} shows the nominal beam parameters requested for this  experiment.

\begin{table}[htbp]
\centering
\begin{tabular}{l|l}
\hline
Momentum {[}GeV/c{]} & 440 \\
Number of bunches  & 288 \\
Bunch intensity {[}protons per bunch{]} & $1.2 \cdot 10^{11}$ \\
Beam distribution & Gaussian \\
Beam size {[}mm{]} & $\sigma_x=\sigma_y= 0.3$ \\ 
Beam pulse {[}$\mu s${]} & $7.2$ \\ \hline
\end{tabular}
\caption{Nominal beam parameters for the HiRadMat facility, where $\sigma$ is the characteristic transverse beam size (in the XY plane).}\label{table2}
\end{table}
The TDI was designed to withstand different kinds of injection kicker  failures. Several kinds of failures of the injector kickers were detected during the TDI´s operation~\cite{lechner2015} and each of them may result in different possible beam orbits and hence impact locations onto the absorbing block. In order to reproduce similar conditions in the experiment, three types of impact were selected:
 \begin{itemize}
\item "Deep" beam impacts, varying the depth of the beam impact into the jaw. The depth is measured with respect to the interface coating-substrate as a multiple of the beam size (represented by $\sigma$).
\item "Grazing" beam impacts, where the centre of the beam is at the interface coating-substrate level (zero $\sigma$ depth) and parallel to the free surface.
\item "Tilted" grazing beam impacts with the beam at a small angle with respect to the coating surface of the block.
\end{itemize}
\subsection{Pre-irradiation examination of the coatings}
\begin{figure}[htbp]
\begin{center}
\subfigure[][]{\includegraphics[width=0.45\linewidth]{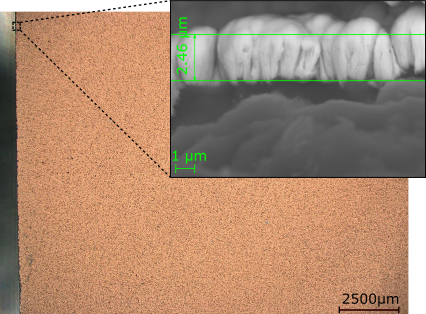}}
\subfigure[][]{\includegraphics[width=0.45\linewidth]{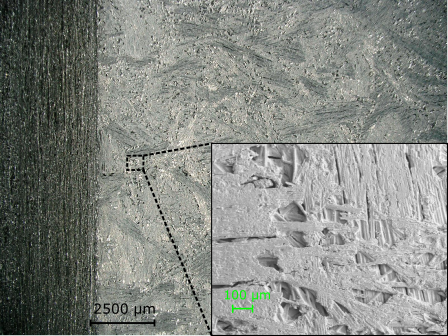}}
\end{center}
  \caption{The figure shows some details of the pre-irradiated appearance of the coatings (top view of the upstream blocks): a) micrographs of the copper coating on the graphite substrate; and b) molybdenum coating on the CfC substrate.}\label{fig3}
\end{figure}

Before irradiation, the initial status of the coating was examined by Optical Microscopy and Scanning Electron Microscopy (SEM) at high resolution. As shown in Fig.~\ref{fig3}, in general, the coated graphite targets presented a satisfactory surface homogeneity with a thickness close to the nominal value $2.5$~$\mu~m$. However, for the CfC-based blocks, although the coating was seen to cover the whole free surface of the blocks, it mimics the fibrous structure of the substrate, resulting in a rough surface finish.

\begin{figure}[htbp]
\begin{center}
\includegraphics[width=0.8\linewidth]{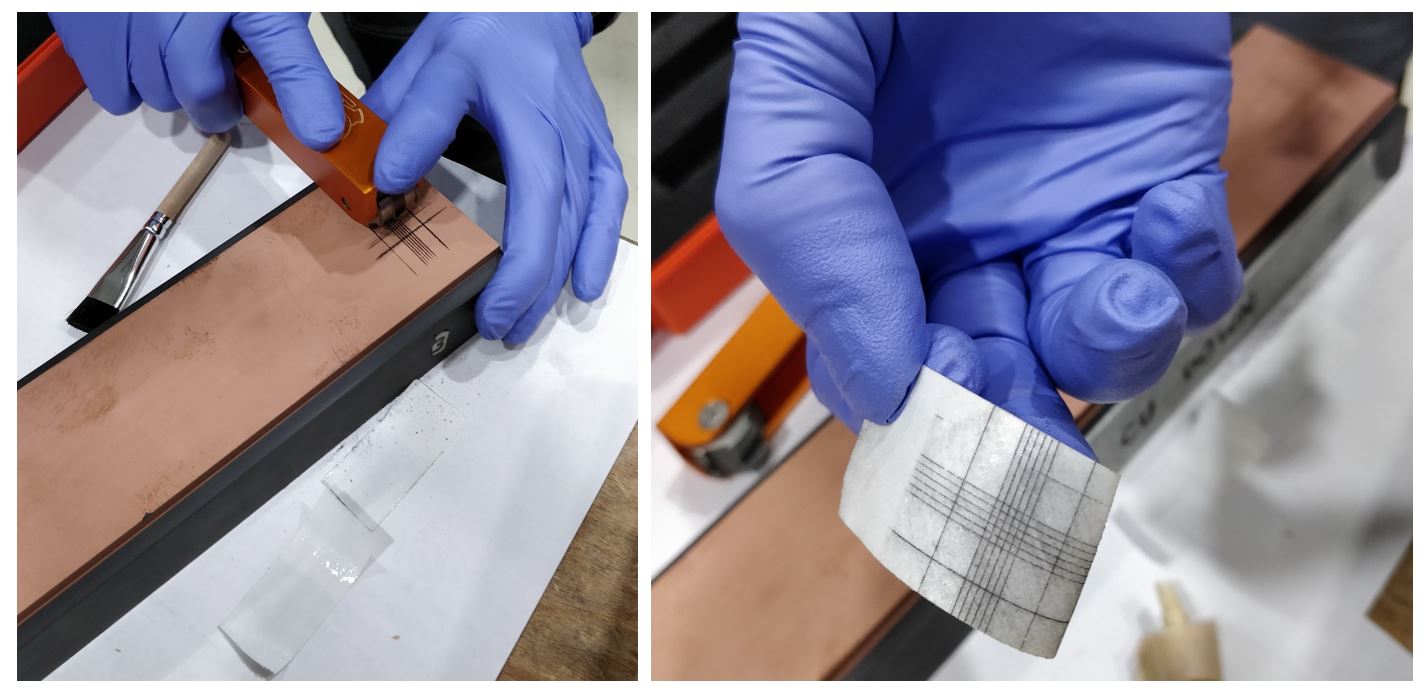}
\end{center}
  \caption{The photo illustrates the tape test on Cu coated graphite blocks, aimed at verifying coating adherence.}\label{fig4}
\end{figure}

Additional tests were performed to verify the quality of the adherence of the coatings to the substrate. A tape test was carried out according to ASTM D3359-B~\cite{d012017standard} using a cross hatch cutter “Elcometer 1542” to make scratches on the coated surfaces. A standardized self-adhesive tape was then used to perform the test; this did not result in any peel-off of any of the coatings. Fig.~\ref{fig4} shows the results of the tape tests on the copper coated graphite sample.  Similar results were found for the molybdenum coated graphite blocks.

\section{Numerical analysis}
\label{Num_Analysis}
The high energy proton beam, used in the HiRadMat experiment, exposes the targets to extremely high thermal loads. The induced rapid temperature increase is followed by a violent mechanical reaction of the material which brings the substrate and coating materials to their structural integrity limits. As part of the design of the present experiment, a series of simulations was carried out to assess the thermo-mechanical response of the targets when impacted with specific beams. The results of these simulations are reported in this section, and help to better understand the post experiment observations described in Section~\ref{Results}). In addition, the numerical analysis findings are used  to deduce some unknown properties of the coatings by comparison with experimental results.

\subsection{Beam-matter interaction simulations}
The interaction of high energetic proton beams with  matter produces hadronic and electromagnetic particle showers, which give rise to a sudden energy deposition in the absorber blocks, coatings and surrounding structures. The FLUKA Monte Carlo software~\cite{Battistoni2015,Bohlen2014,FlukaWeb} was used for the calculation of the energy deposition. In all cases, we assumed that the density of the copper and molybdenum coatings corresponds to the nominal bulk density of these materials.

\begin{figure}[htbp]
\begin{center}
{\includegraphics[width=0.3\linewidth]{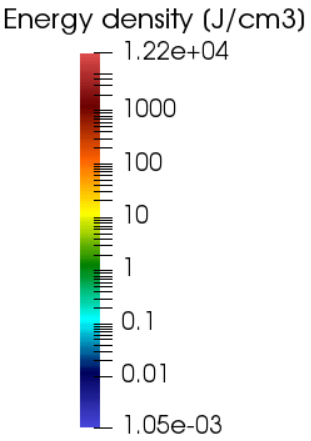}}
{\includegraphics[width=0.64\linewidth]{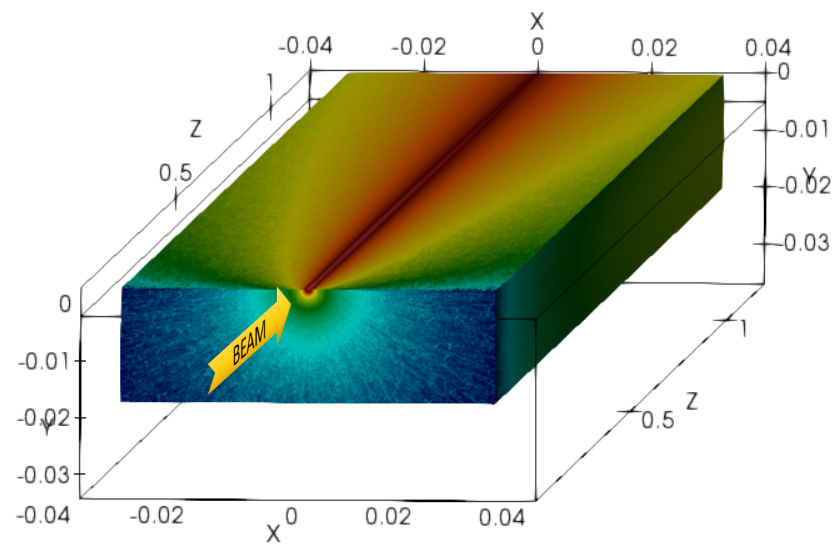}}
\end{center}
 \caption{The image shows an example of the FLUKA-generated~\cite{Battistoni2015,Bohlen2014,FlukaWeb} energy density deposited in the Cu coated graphite block under a grazing beam impact.}\label{fig5}
\end{figure}

Fig.~\ref{fig5} shows the 3-D energy deposition map for the copper-coated graphite block subjected to a grazing beam impact. Two different meshes were put in place around the incidence region to quantify the energy deposited in the coatings and substrates, respectively. A homogeneous coating with a nominal thickness of $2.5$~$\mu m$, was considered. A mesh size of $\sigma/4$ was set in the plane transversely to the beam with at least one element through the coating thickness, and $5$~mm along the beam direction, in order to provide a suitable spatial resolution.

It is important to notice that the energy deposition density in the coating is notably higher than in the substrate mainly due to its higher material density. This phenomenon has implications for the thermal evolution of the coating, as shown in Sec.~\ref{Thermal_Analysis_Coating}.

\begin{figure}[htbp]
\begin{center}
\begin{tabular}{p{0.35\textwidth} p{0.6\textwidth}}
  \vspace{0pt} \includegraphics[width=0.35\textwidth]{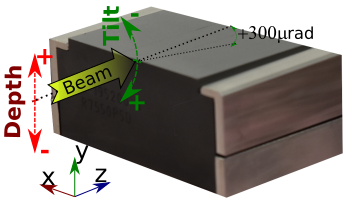} &
  \vspace{0pt} \includegraphics[width=0.59\textwidth]{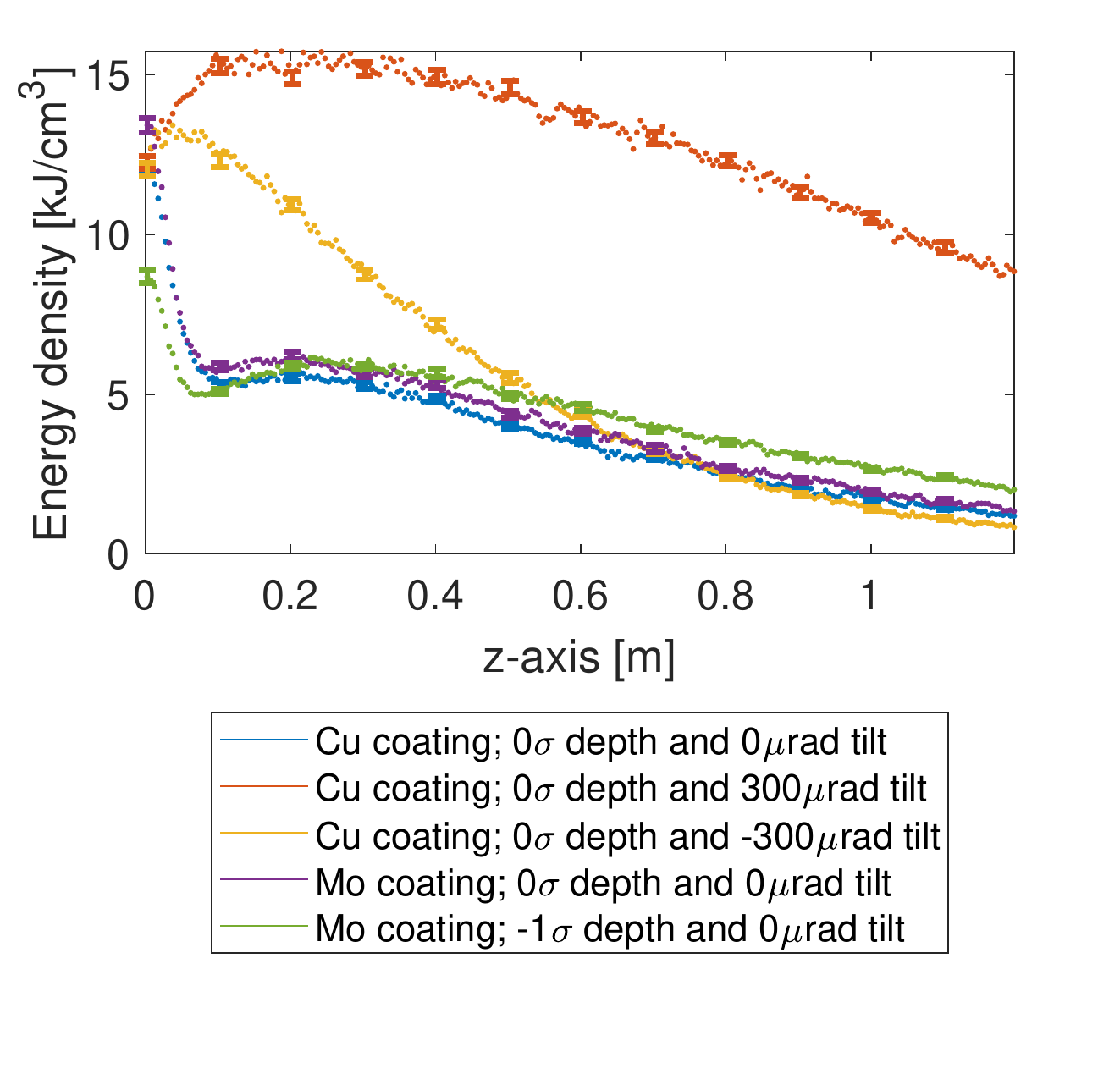}
\end{tabular}
\end{center}
 \caption{Energy density deposited in the copper and molybdenum coatings along the beam direction (z-axis) for two different beam impact depths ($0\sigma$ and $-1\sigma$) and beam angles ($300$~$\mu~rad$ and $-300$~$\mu~rad$). Results are shown for the graphite substrate. The relevant reference system is shown on the left picture where the beam trajectory is defined with respect to the interface coating-substrate and given by the depth and tilt angle as shown by the arrows.}\label{fig6}
\end{figure}

Fig.~\ref{fig6} shows the energy density deposited in the copper and molybdenum coatings along the beam direction for different beam impact depths and tilt angles with respect to the coating surface (note that the reference system used along the paper is included in the figure). Both coatings exhibit similar energy deposition patterns. The peak of the energy deposition density is found at the entrance region for a grazing beam impact. Deeper impacts in the target lead to a reduction of peak energy density. The highest levels of energy deposition density are found for positive tilts (as defined in Fig.~\ref{fig6}), and the energy density peak is shifted downstream. Tilted grazing impacts are the most demanding  ones for the coatings from the energy deposition point of view.

\begin{figure}[htbp]
\begin{center}
\subfigure[][]{\includegraphics[width=0.49\linewidth]{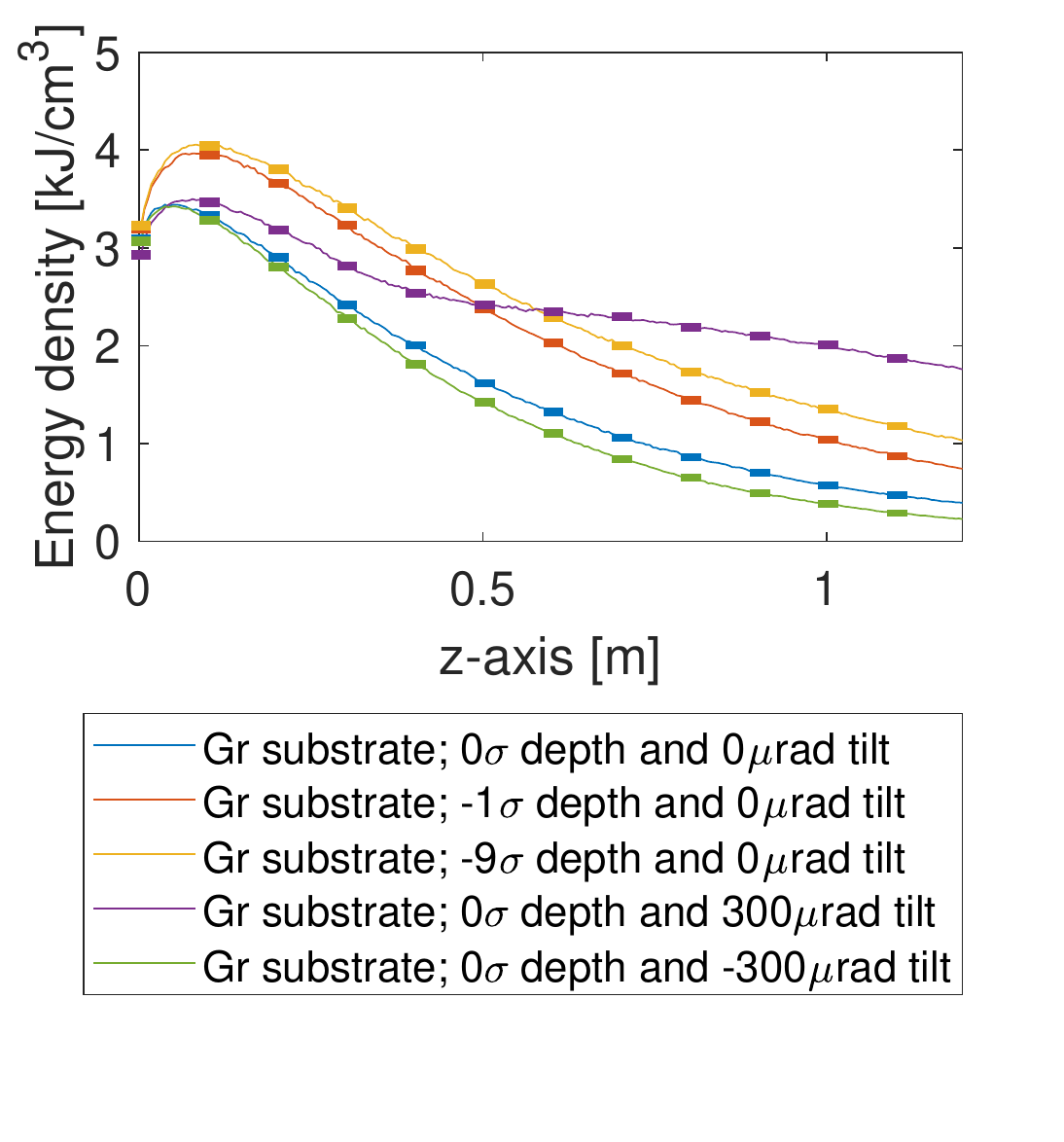}}
\subfigure[][]{\includegraphics[width=0.49\linewidth]{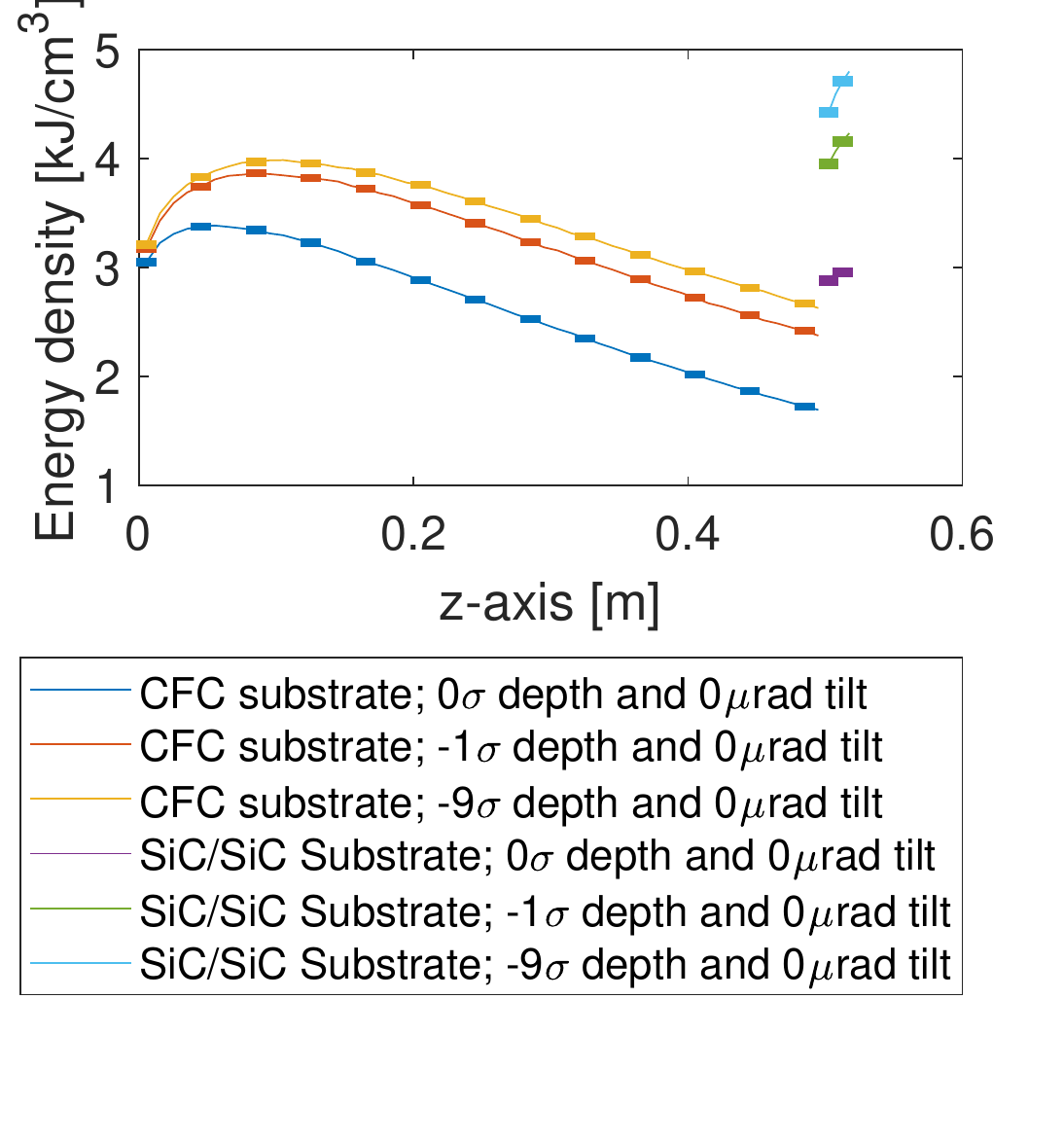}}
\end{center}
 \caption{Energy density deposited in the a) graphite and b) CfC and SiC/SiC substrates along the beam direction for different beam impact depths ($0\sigma$ and $-1\sigma$) and beam angles ($300$~$\mu rad$ and $-300$~$\mu rad$). Note that the SiC/SiC block was installed downstream of the CfC block.}\label{fig7}
\end{figure}

The energy density deposited in the graphite, CfC and SiC-SiC substrates is shown in Fig~\ref{fig7}. The energy deposition increases notably with the depth, which makes it potentially more harmful. The beam tilt changes the energy deposition pattern, but the peak energy density is not significantly modified in the substrate.

\subsection{Thermal analysis of the coatings}
\label{Thermal_Analysis_Coating}

\begin{figure}[htbp]
\begin{center}
\includegraphics[width=0.55\linewidth]{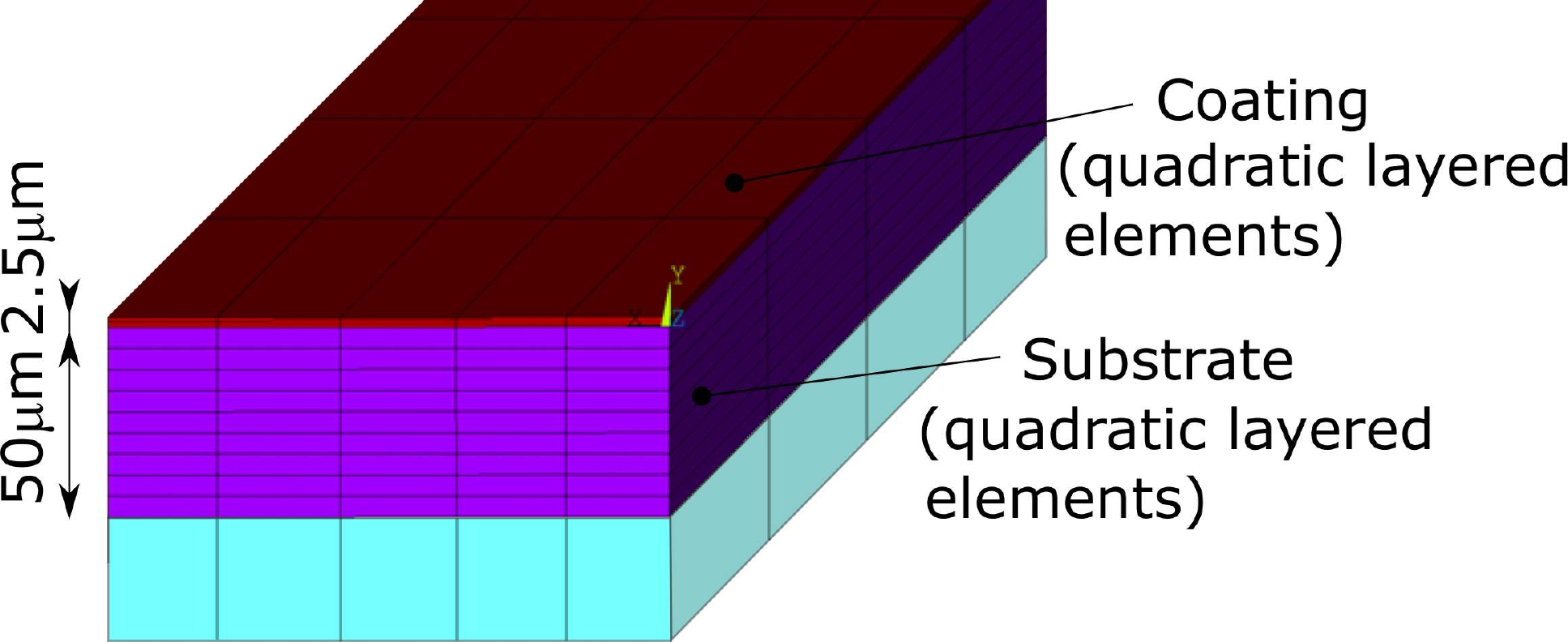}
\end{center}
 \caption{Thermal FEM model of the coated substrates based on a layered formulation for an efficient and accurate simulation of the characteristic thermal diffusion field.}\label{fig8}
\end{figure}

When considering the energy deposition in the target, adiabatic models are usually valid for the estimation of the peak temperature in fast thermal events as studied in the present work. The material does not have time to diffuse the thermal energy and mechanical waves propagate much faster than thermal energy. Nevertheless, in this particular case of coated absorbing targets, the heat transfer phenomenon plays an important role in the temperature distribution and evolution of the coating, even in only $7.2$~$\mu$s  beam impact duration. For this reason, a dedicated thermal Finite Element Method (FEM) model, using ANSYS\textsuperscript{\textregistered} thermal was developed to predict the temperature evolution in the coatings (see Fig.~\ref{fig8}).

The model was restricted to the graphite substrate that has a homogeneous coating layer. Shell layered elements (shell131) with a quadratic basis function along the thickness were used to accurately represent the coating, whilst solid brick elements (solid70) were used for the substrates. Note that the characteristic length of the thermal diffusion for the graphite is $\mu_{t=7.2\mu s}^{Gr}=47~\mu m$, which provides an estimation of the thermal diffusion affected area during the beam pulse. Layered elements were also applied along the first $50$~$\mu$m depth of the substrate to accurately capture this phenomenon. This technique allows a reduction of the number of Degrees of Freedom and elements without impacting the accuracy because the variation of temperature inside the element is addressed internally at the level of the integration points. This approach also provides a better element aspect ratio and enhances numerical stability. Bulk properties of the copper and molybdenum coating were considered \cite{ditmars1977,Valencia2010}. The melting process was included in the model by considering the change of the thermo-physical properties during the phase change. Mesh and time-step independent studies were carried out to set up the numerical model.

The main parameter that determines the temperature in the coating is the Thermal Contact Conductance (TCC) between the coatings and the substrates. The TCC depends on many variables, such as the material properties, surface roughness, coating adherence, etc, and hence it is closely related to the coating quality. This parameter is a-priori unknown, but it can be inferred by  cross-checking and comparing the experimental results with the numerical model findings. This approach can provide some reference values that could guide future studies on coated absorbing materials.

\begin{figure}[htbp]
\begin{center}
\subfigure[][]{\includegraphics[width=0.7\linewidth]{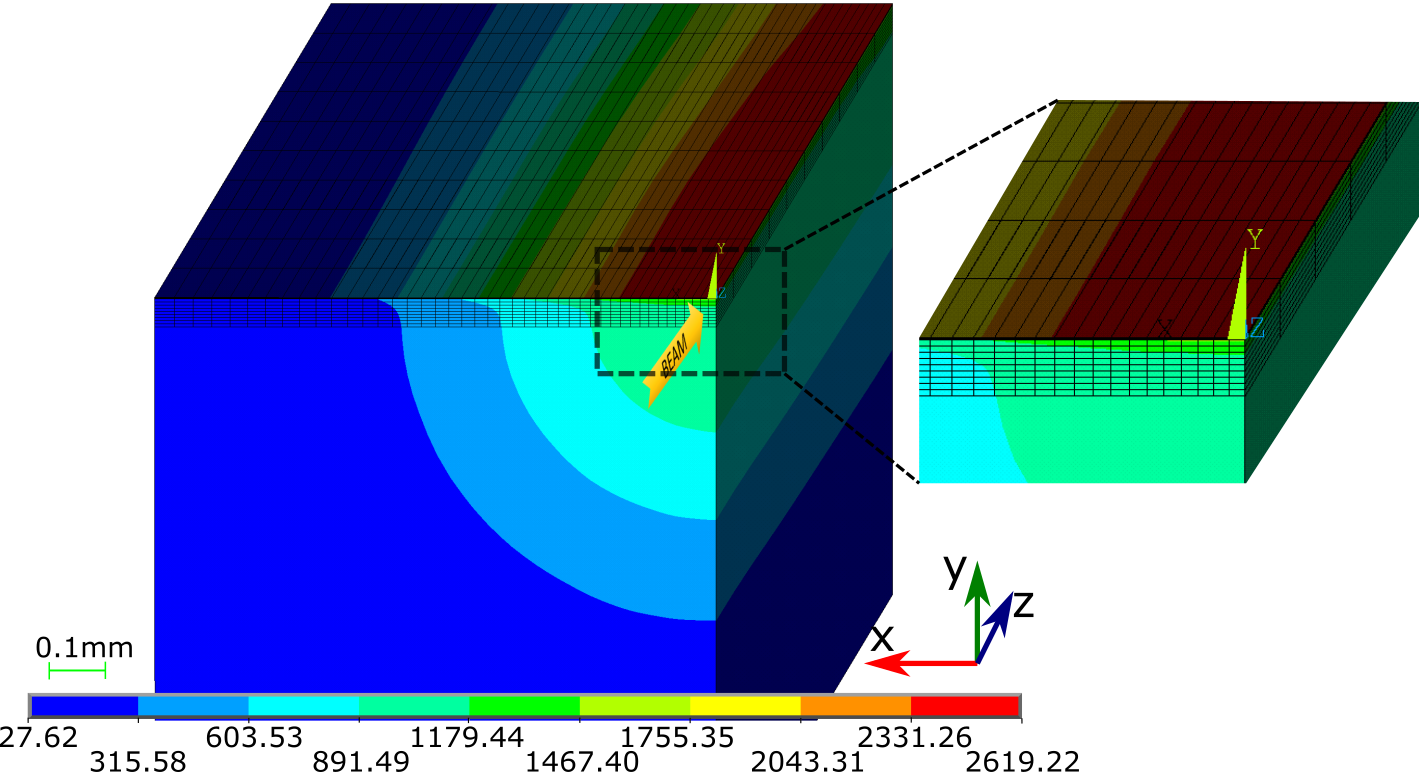}}
\subfigure[][]{\includegraphics[width=0.55\linewidth]{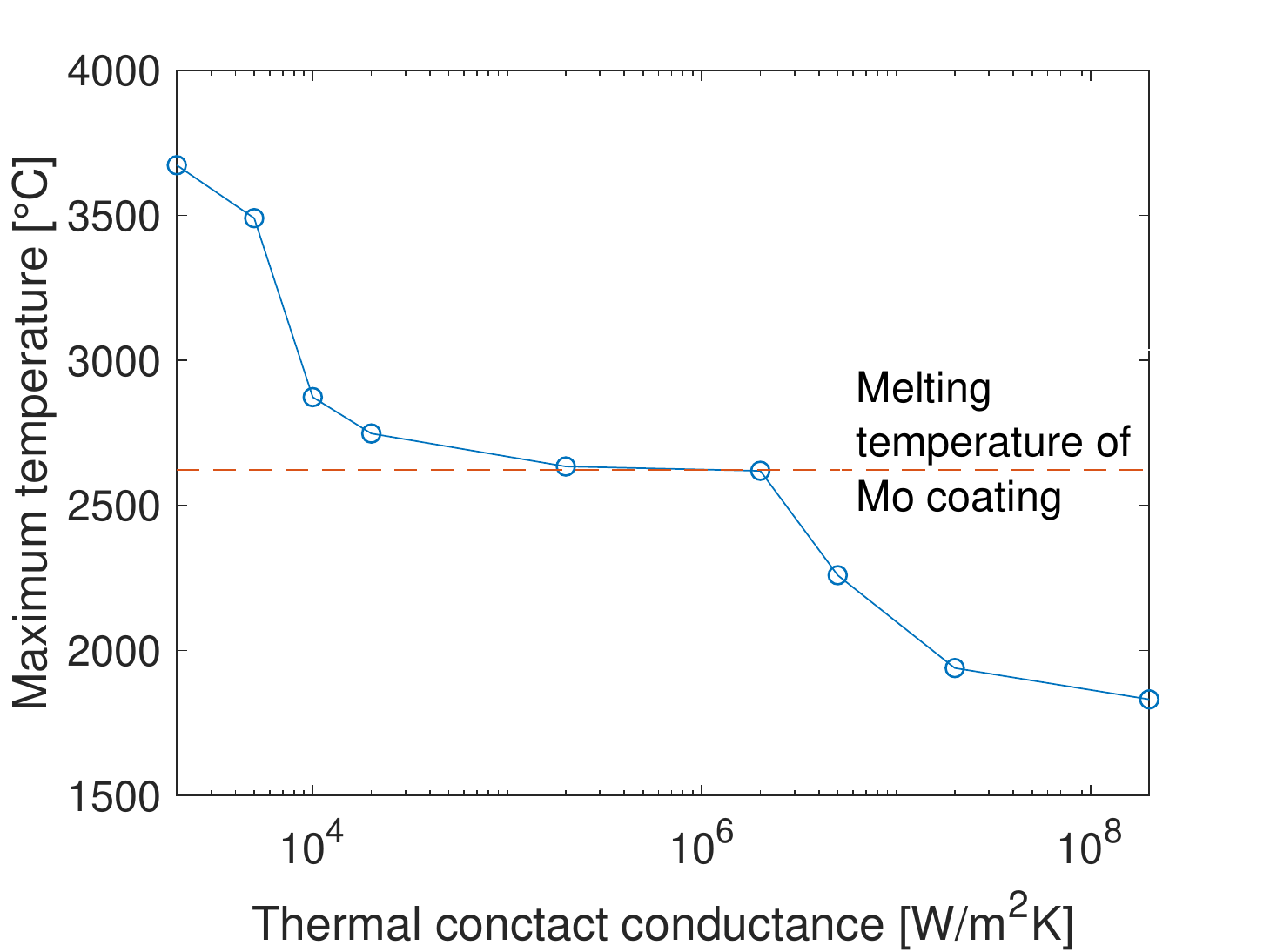}}
\end{center}
 \caption{a) Temperature distribution in the molybdenum-coated graphite block after a grazing beam impact for a $TCC_{Mo-Gr}^*=2.00\cdot 10^6$ $[W/m^2K]$ (half model considering YZ symmetry plane). b) Maximum temperature in the molybdenum coating under a grazing beam impact versus the thermal contact conductance between the coating and graphite substrate}\label{fig9}
\end{figure}

Fig.~\ref{fig9}a shows the temperature distribution in the molybdenum-coated graphite block after the beam pulse ($t=7.2$~$\mu s$) assuming a given TCC. The FLUKA-derived energy density map was imported as the heat load input. The temperature in the coating is noticeably higher than in the graphite substrate mainly due to  the greater deposited energy density and lower specific heat capacity of the coating. The temperature gradient across the first $50$~$\mu$ m of the substrate next to the coating is relatively high, which illustrates the importance of the thermal diffusion from the coating to the substrate. 

The peak temperature  in the molybdenum coating versus the TCC is shown in Fig.~\ref{fig9}b. The peak temperature decreases with the TCC and the coating starts to melt at the beam entrance point for values lower than $TCC_{Mo-Gr}^*=2.00\cdot 10^6 [W/m^2K]$, which demonstrates the importance of this parameter. A similar phenomenon can be observed in the copper coating.

\subsection{Thermo-mechanical analysis of the substrates}
\label{FEA_substrate}

The mechanical behaviour of the substrates was analysed by a multi-physic FEM (LS-DYNA\textsuperscript{\textregistered} software) in order to assess the stress induced by the beam impact. A continuous and homogeneous medium was considered for the different substrates, assuming an isotropic elastic model for the isostatic graphite and an orthotropic model for the CfC. Both materials are frequently used in beam intercepting devices at CERN and they have been widely characterized~\cite{guradia2018,bianchi2017,Sacristan2017}.

The model was focused on the first $180$~mm  length from the upstream end of the blocks, corresponding to the area where the maximum energy deposition in the substrate occurs. Non-reflective boundary conditions were applied to the virtual cut face in order to avoid unrealistic wave reflections. A mesh refinement around the beam impact was implemented in order to capture the temperature and stress gradient produced by the localized beam energy deposition, with a characteristic size of $L_{ch}=1/20 \cdot \sigma$ (linear elements). A mesh sensitivity analysis was carried out to avoid any mesh influence.

\begin{figure}[htbp]
\begin{center}
\subfigure[][]{\includegraphics[width=0.65\linewidth]{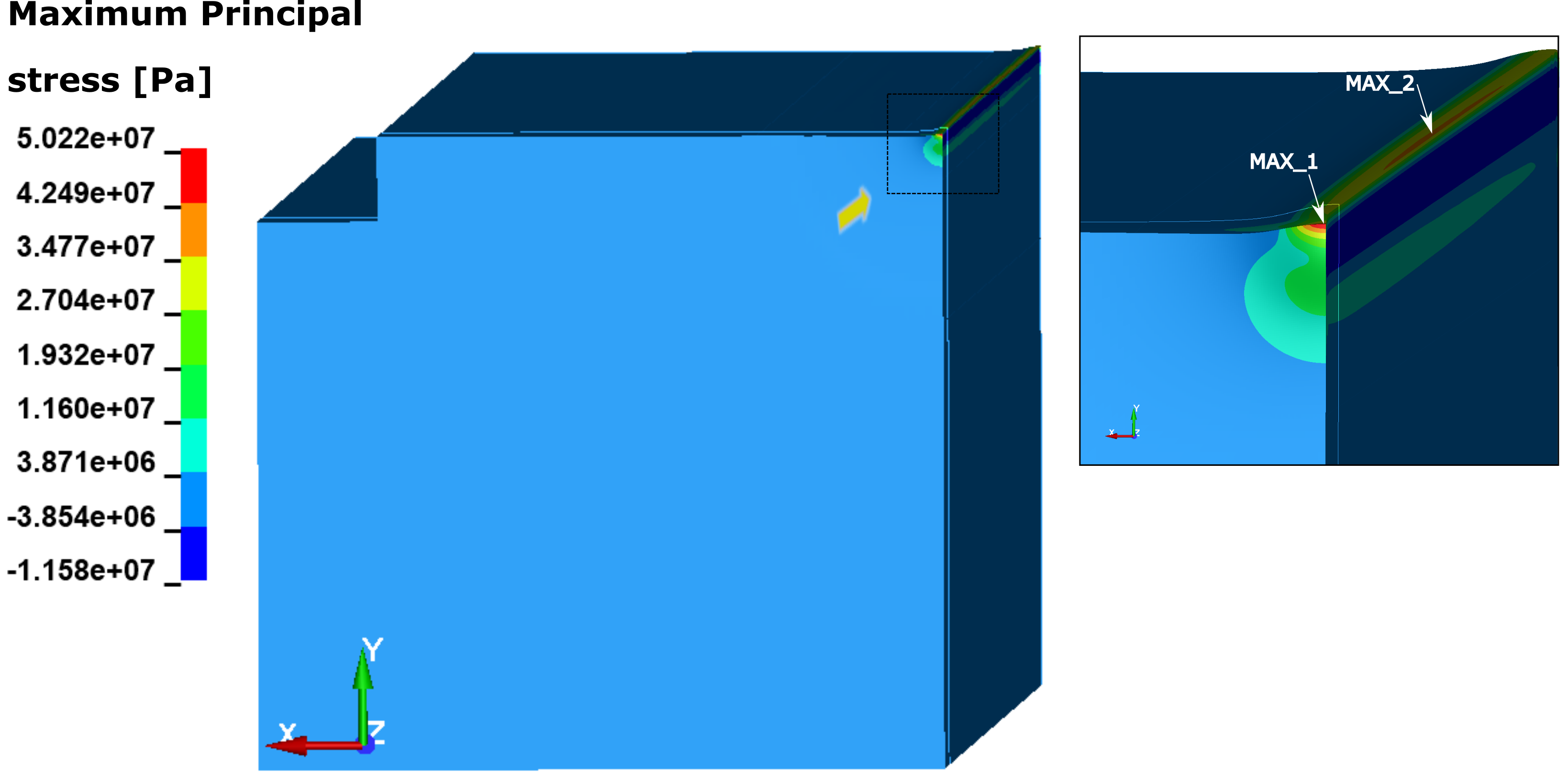}}
\subfigure[][]{\includegraphics[width=0.4\linewidth]{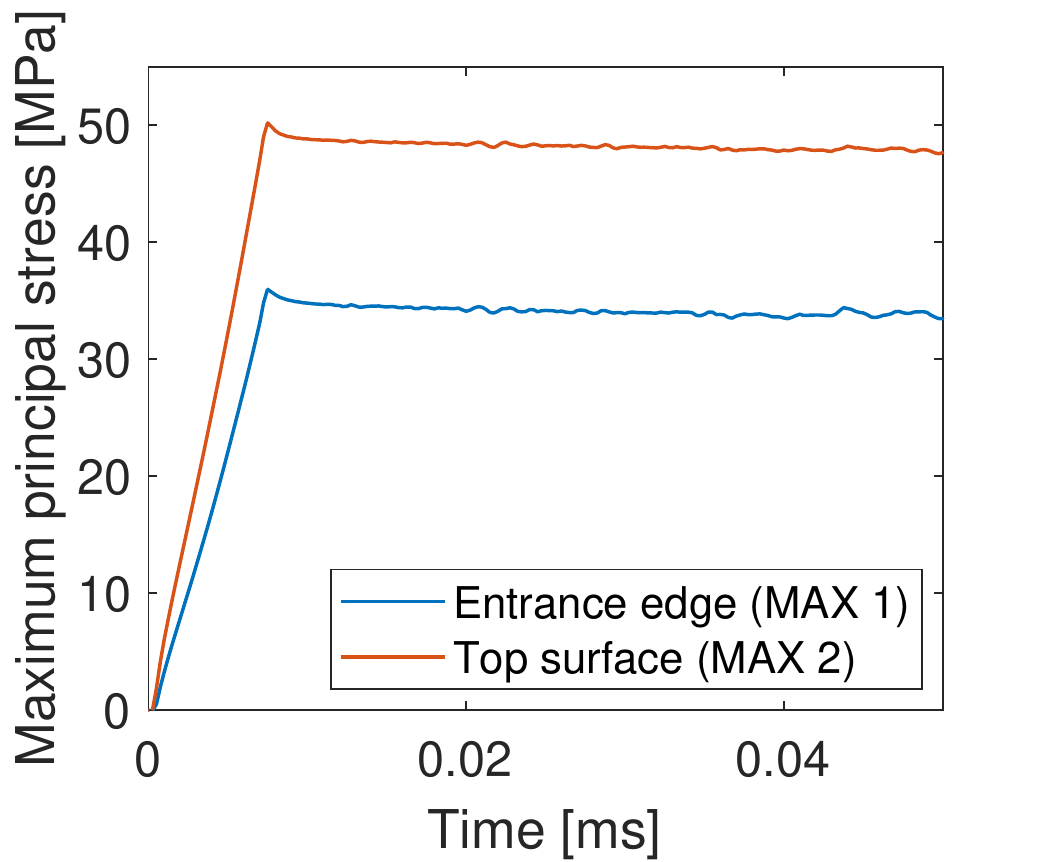}}
\subfigure[][]{\includegraphics[width=0.4\linewidth]{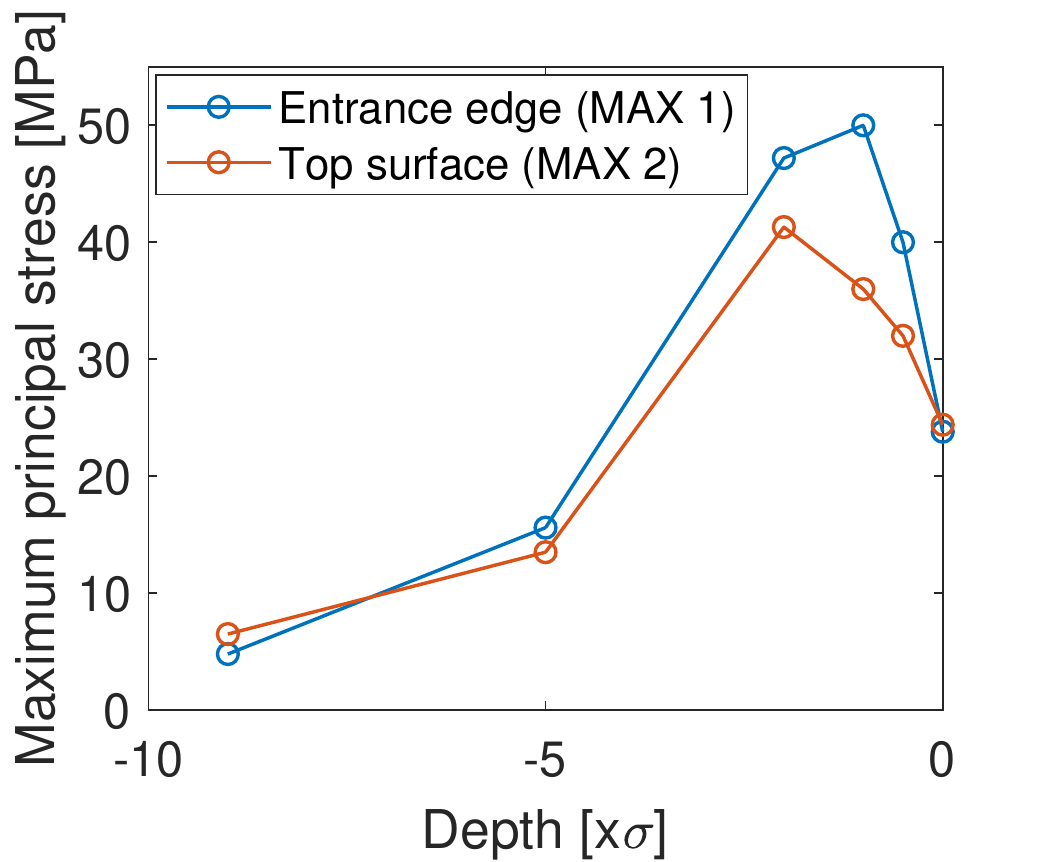}}
\end{center}
 \caption{a) Maximum principal stress in the graphite substrate after the beam pulse (at $t=7.5\cdot 10^{-6}$ $\mu$s) for a deep impact at $-1\sigma$. Only half of the block is  represented considering the YZ symmetric plane. Arrows point to zones with the maximum stresses. Displacements have been amplified by a factor 50. b) Maximum stress evolution in the most critical points over time. c) Peak of maximum principal stress for different beam impact depths.}\label{fig10}
\end{figure}

The maximum stresses were found for the deep impact type at the entrance face and along the free horizontal surface where the beam passes in parallel (top surface). Fig.~\ref{fig10}a shows the maximum stress in the graphite substrate for a deep impact at $-1\sigma$. After the impact of the proton beam, the pressure suddenly increases around the beam collision region, the elastic waves propagate into the substrate and the free surfaces (top, entrance and exit surfaces) deform, producing a bump, these surfaces being the most stressed.

Although the load is inherently dynamic, stresses at the free surfaces become quasi-static after the beam pulse (see Fig.~\ref{fig10}b). Moreover, it is observed that even if a deeper impact ($<-1\sigma$) is potentially more harmful in terms of deposited energy, the stresses at the free surface are lower due to its increased distance to the beam impact location (Fig.~\ref{fig10}c). 

\begin{figure}[htbp]
\begin{center}
{\includegraphics[width=0.55\linewidth]{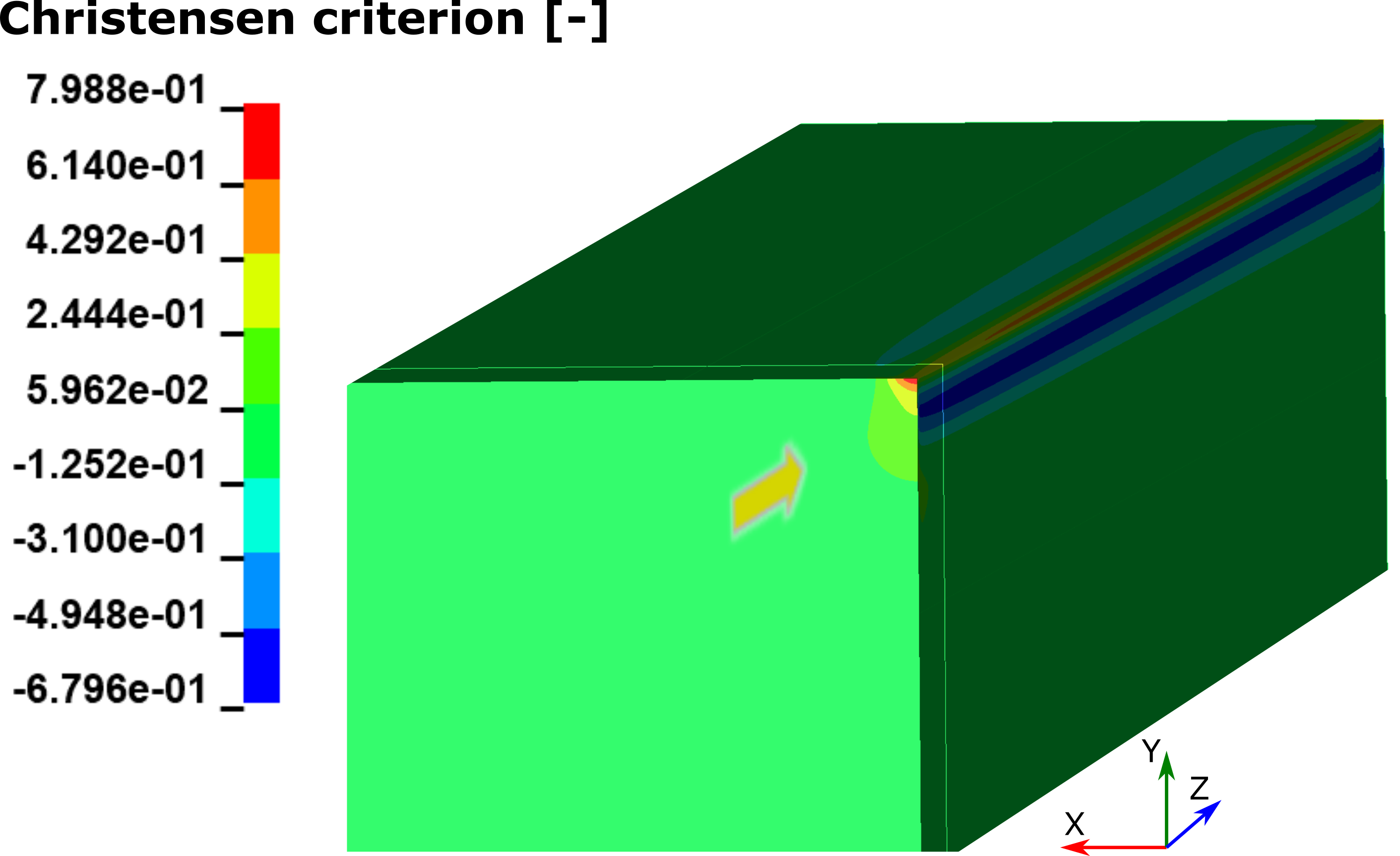}}
\end{center}
 \caption{Christensen failure criterion for brittle materials applied to the graphite substrate for a deep impact ($-1\sigma$). The maximum value is $0.8$ on the top surface (coating surface), where values greater than 1 mean failure. Note that only half of the block is represented considering the YZ symmetric plane.}\label{fig11}
\end{figure}

Fig.~\ref{fig11} shows the equivalent dimensionless Christensen failure criterion~\cite{Christ2007}, typically applied in brittle materials, in the graphite substrate for a deep impact at $-1\sigma$. No damage is expected since the maximum Christensen's value is $0.8$ and values greater than one would imply damage. Similar analyses were carried out for the CfC substrate, resulting in  larger safety margins than for graphite (inverse reserve factor $IRF_{max}=0.55$  based on the maximum stress criterion, where values greater than 1 mean failure).

Given the particular structure of the SiC-SiC block, made of unidirectional alternating layers at 0/90 degrees, a three-dimensional FEM model based on the layered technology at the level of each ply was developed. A detailed mechanical characterization, including  failure analysis of this particular SiC-SiC composite, can be found in Ref.~\cite{nozawa2012,yb2012} and is used as baseline for the present study.

\begin{figure}[htbp]
\begin{center}
\includegraphics[width=0.5\linewidth]{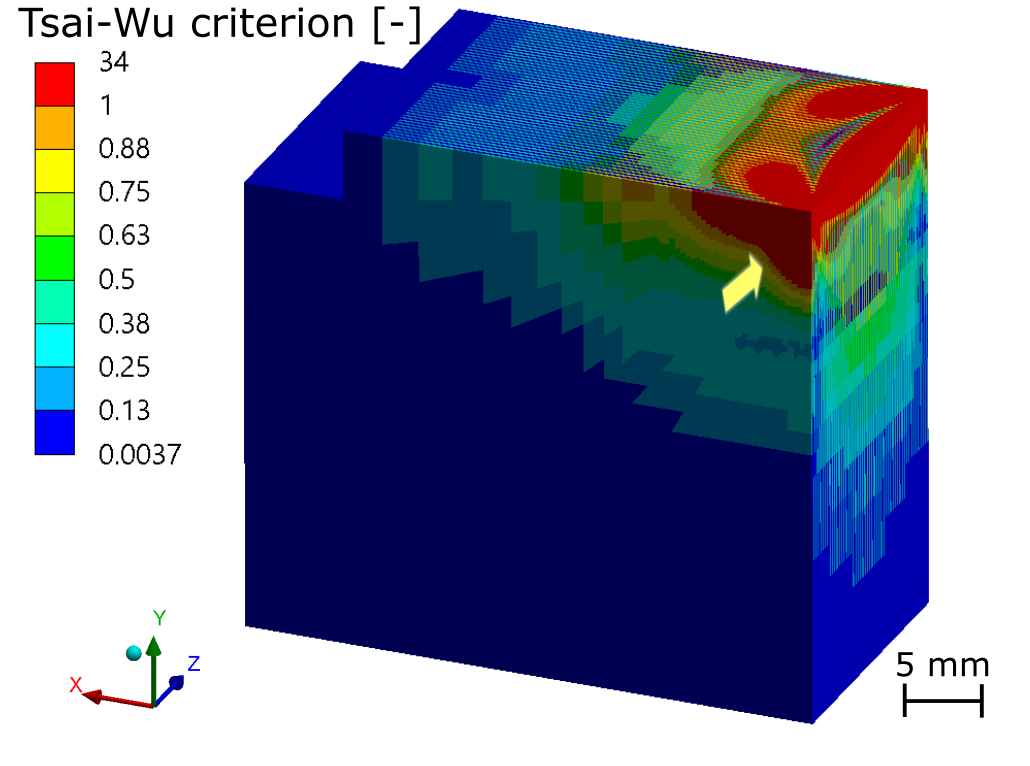}
\end{center}
 \caption{Tsai-Wu failure criterion on the SiC-SiC after beam impact for a deep impact at $-1\sigma$ (half symmetric model). Values greater than 1 mean failure. Note that only half of the block is represented considering the YZ symmetric plane.}\label{fig12}
\end{figure}

As suggested in Ref.~\cite{nozawa2012}, and typically used in unidirectional layered composites~\cite{LI2017207}, the Tsai-Wu criterion~\cite{tsai1971} was used to predict damage in the SiC-SiC block. This criterion, based on a tensorial stress function, it is able to takes into account the orthotropic nature of this material. As shown in Fig.~\ref{fig12}, in the case of a deep impact (the most severe type), damage is predicted all along the free surface parallel to the beam impact, where the entrance and exit faces are the most affected areas. These results were compared with the experimental observations and are discussed in Section~\ref{Substrate_analysis}).

%\section{Experimental execution}
\section{Experiment with proton beams, instrumentation and measurements}

\subsection{Proton beams used in the experiment}
\label{Exp_execuation}
The jaws were irradiated with up to 23 proton beams at nominal intensity, for each of the three types of beam impacts (see Table~\ref{tab_2}). Each type of beam impact (deep, grazing and tilted impact) was horizontally offset in the jaw by more than $23 \sigma$ to make the post-irradiation analysis of the respective beam-induced phenomena easier and avoid effects from the other unrelated beam impacts.

\subsection{Main instrumentation and measurements}
\begin{figure}[htbp]
\begin{center}
\includegraphics[width=0.65\linewidth]{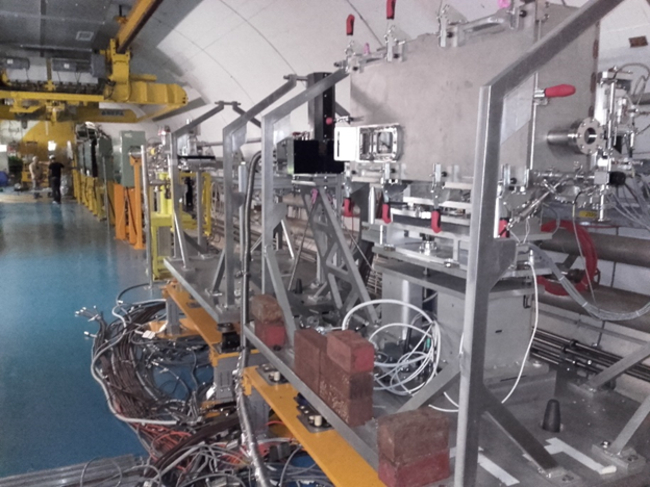}
\end{center}
 \caption{Closed tank integrated in its handling module and installed in the HiRadMat experimental zone.}\label{fig13}
\end{figure}

Figure~\ref{fig13} shows the final integration of the tank in the beam line of the HiRadMat facility. A precise knowledge and control of the beam size and position relative to the targets during the experiment was required in order to fully separate the different types of impacts and discern their effects. To obtain this information, the experiment was equipped with a set of instruments consisting of: 
\begin{itemize}
\item One beam position monitor (BPKG)~\cite{Forck2009}, placed just upstream of the jaws in order to accurately monitor the beam position.
\item Two Beam Loss Monitors (BLMs)~\cite{zhukov2010} to monitor beam losses during beam alignment and grazing impact.
\item One Beam Observation TV Monitor (BTV)~\cite{Bravin2005} located upstream of the jaws to precisely measure the beam size.
\item Seven Linear Variable Differential Transformer (LVDT) sensors to measure the positions and displacements of the different jaws.
\end{itemize}

\begin{figure}[htbp]
\begin{center}
\includegraphics[width=0.99\linewidth]{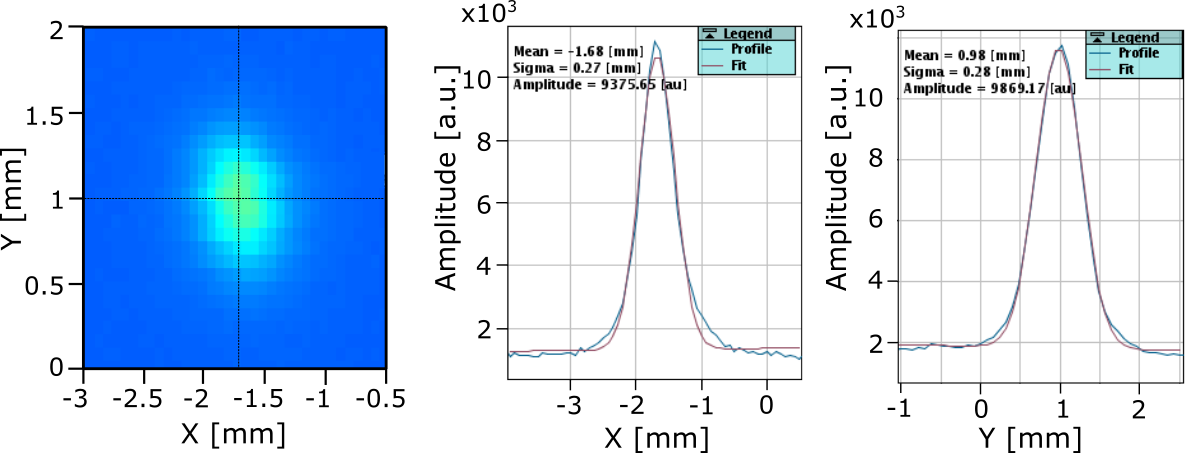}
\end{center}
 \caption{BTV image of one of the pilot beams (made by one single bunch with $1.2\cdot 10^{11}$~protons/bunch) used during the beam-based alignment of the Mo coated CfC jaw.}\label{fig14}
\end{figure}

The BTV is based on a flat screen that interacts with the proton beam to generate light, thereby showing accurately the beam profile. Fig.~\ref{fig14} shows the profile for one pilot beam, which exhibits a bi-Gaussian shape. During the experiment, the beam size could be correctly approximated to the nominal value of $\sigma=0.3$~mm.

\begin{figure}[htbp]
\begin{center}
\subfigure[][]{\includegraphics[width=0.49\linewidth]{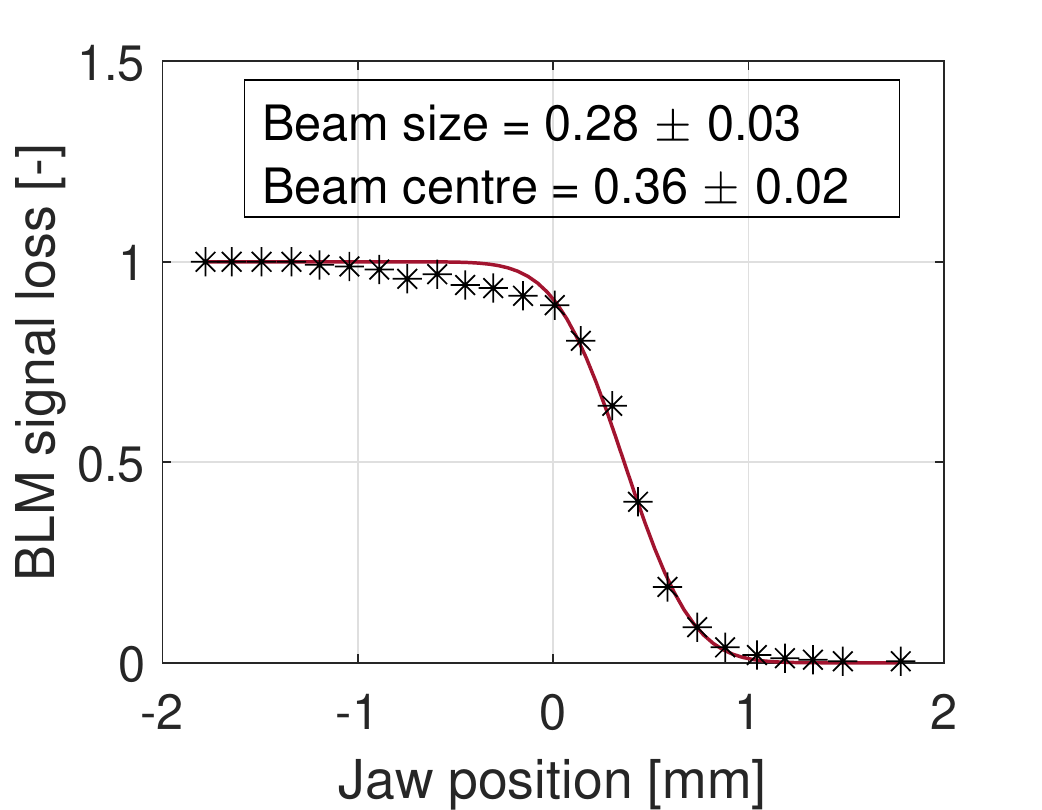}}
\subfigure[][]{\includegraphics[width=0.49\linewidth]{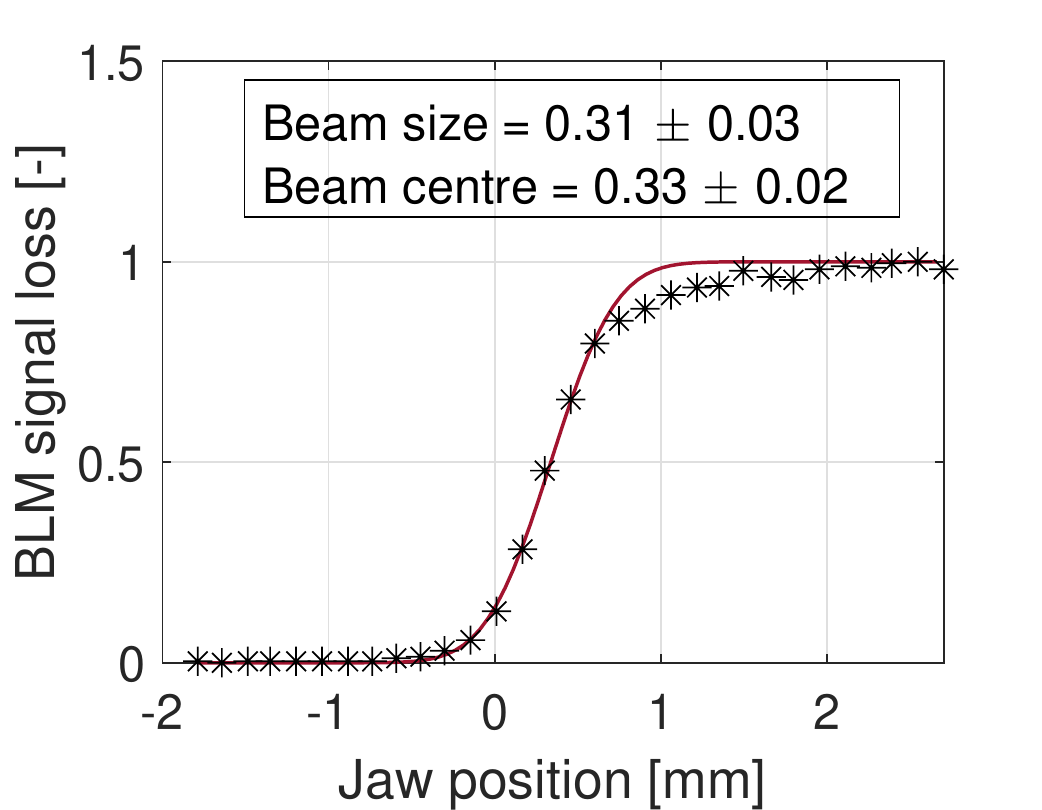}}
\end{center}
 \caption{Experimental beam parameters obtained from beam-based alignment measured on the a) top jaw (Mo coated CfC jaw) b) bottom jaw (Cu coated graphite jaw). Beam losses have been normalized in the range of $[0-1]$. Note that the crosses correspond with the measurements and red line with the fitting Gaussian error function curve.}\label{fig15}
\end{figure}

The exact position of the beam with respect to the jaw was set-up by a beam-based alignment procedure relying on BLM measurements over a series of individual beam impacts with pilot beams at low intensity. The BLM measures the secondary particle showers that result from the beam-jaw interaction. Beam losses were measured for different relative positions of the coating surface of the jaw with respect to the beam trajectory. Assuming a Gaussian beam profile, the discrete series of measurements, which relate the induced beam losses with the absolute position of the jaw, can be fitted by a Gaussian error function ($erf(\bar{y})$, with $\bar{y}=(y-y_0)/(\sqrt{2}\sigma)$ where $y$ is the jaw position, $y_0$ is the beam centre and $\sigma$ is the beam size). Therefore, the absolute position of the beam centre with respect to the jaws can be directly inferred with high accuracy (see Fig.~\ref{fig15}).

\subsection{Additional instrumentation}
%The experiment was executed in primary vacuum conditions ($10^{-3}$ $mbar$) to prevent oxidation of the carbon blocks at high temperature. 
The experiment was monitored live by means of five radiation resistant cameras focused on the jaws. Radiation-hard temperature sensors (Pt100) were also installed; these indicated that the time between consecutive impacts was enough to allow the cool down of the jaw  to practically room temperature.

\begin{table}[htpb]
\begin{center}
%\begin{adjustbox}{width=0.49\textwidth}
\begin{tabular}{llcc}
Jaw & \begin{tabular}[c]{@{}l@{}}Type \\ of impact\end{tabular} & \begin{tabular}[c]{@{}c@{}}Beam impacts \\ {[}Quantity{]}\end{tabular} & \begin{tabular}[c]{@{}c@{}}Avg. intensity\\  {[}$10^{13}$\ protons{]}\end{tabular} \\ \hline
\multicolumn{1}{l|}{\multirow{3}{*}{\begin{tabular}[c]{@{}l@{}}Cu-coated \\ graphite\end{tabular}}} & Deep & 10 & $3.56$\\
\multicolumn{1}{l|}{} & Grazing & 6 & $3.56$\\
\multicolumn{1}{l|}{} & Tilted & 6 & $3.53$\\ \hline
\multicolumn{1}{l|}{\multirow{3}{*}{\begin{tabular}[c]{@{}l@{}}Mo-coated \\ graphite\end{tabular}}} & Deep & 11 & $3.48$\\
\multicolumn{1}{l|}{} & Grazing & 6 & $3.48$\\
\multicolumn{1}{l|}{} & Tilted & 6 & $3.45$\\ \hline
\multicolumn{1}{l|}{\multirow{3}{*}{\begin{tabular}[c]{@{}l@{}}Mo-coated CfC \\ + SiC\textbackslash{}SiC\end{tabular}}} & Deep & 7 & $3.54$\\
\multicolumn{1}{l|}{} & Grazing & 6 & $3.55$\\
\multicolumn{1}{l|}{} & Tilted & 6 & $3.54$
\end{tabular}
%\end{adjustbox}
\end{center}
\caption{Experimental summary of beam impacts on each jaw and its targets.}\label{tab_2}
\end{table}

\section{Experimental results and discussion} 
\label{Results}

Several months after the beam impact operations in HiRadMat (allowing the samples to reduce their residual dose rate levels to values compatible with handling in the laboratory), a post irradiation examination (PIE) campaign was launched to investigate the performance of the different targets under beam impacts. This campaign included: Scanning Electron Microscopy (SEM), Optical Microscopy, tomography and impedance measurements. The results are discussed in this Section and compared with the numerical analysis findings.
\subsection{Analysis of the coatings}
\subsubsection{Copper coating}

\begin{figure}[htbp]
\begin{center}
\includegraphics[width=0.8\linewidth]{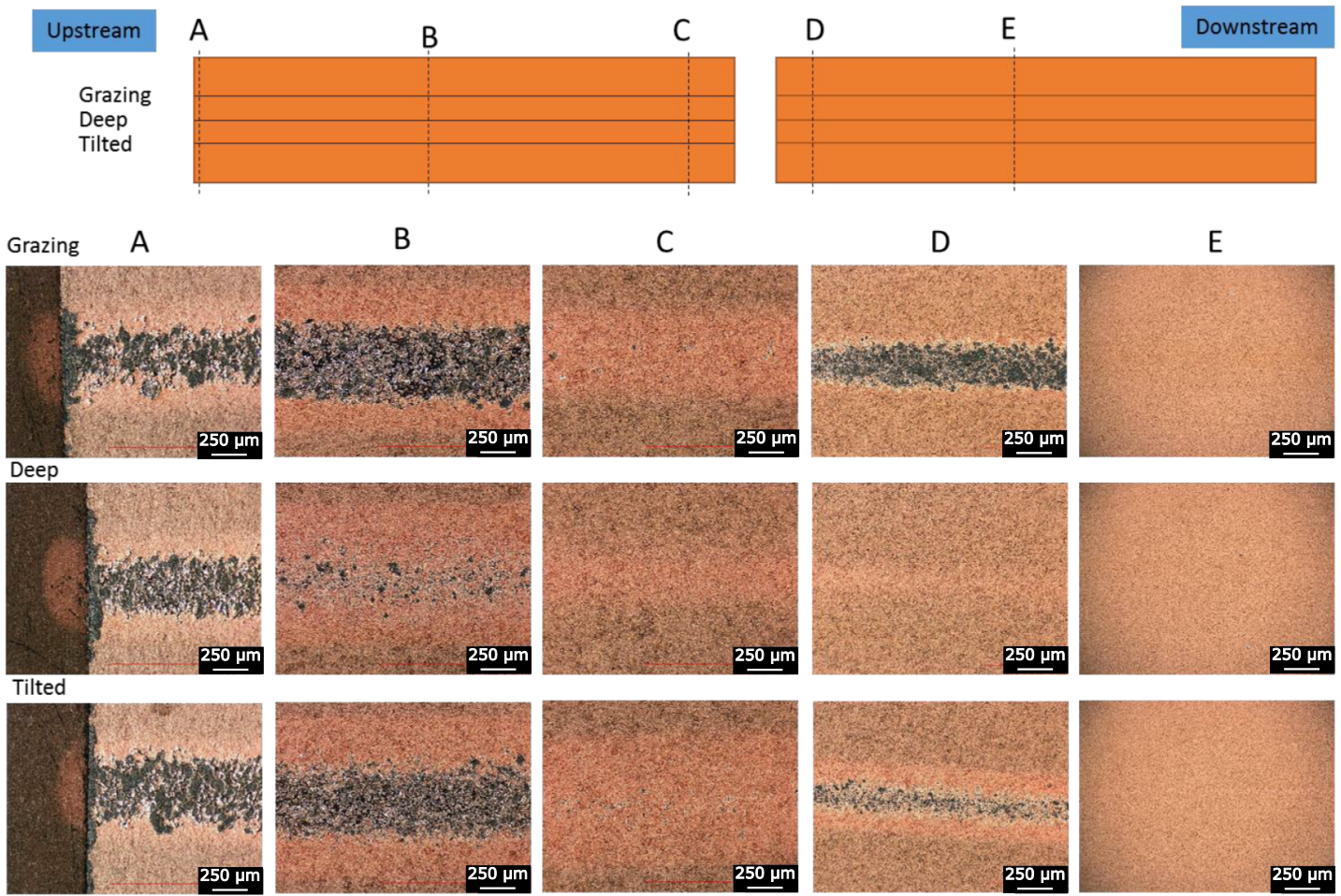}
\end{center}
 \caption{Optical microscope images of the Cu-coated graphite jaw acquired at different points along the beam impact axes, showing an overview of the coating damage for the three different kinds of impact. Note that the positions along the blocks where the images were taken is shown in the schematic illustration at the top of the figure.}\label{fig16}
\end{figure}

An initial visual inspection revealed that the coatings were clearly damaged. Optical microscope images of the copper coated graphite jaw were acquired at several areas along the three beam impact locations. Fig.~\ref{fig16} provides a full overview the jaw. 

The coating damage was observed only in the upstream areas of both copper coated blocks in the jaw. The coating appears to have been scraped away and the graphite substrate has become partially visible. The coating damage is more pronounced for the grazing and tilted impacts. For the deep impact case, the damage is mainly localized at the entrance region. These results are coherent with the energy deposition studies showing that the deposited energy in the coating decreases with the impact depth and position along the jaw axis. The damage observed the beginning of the second block may suggest a possible slight misalignment between both blocks, that may induce an over-exposure of the coating of the second block.

\begin{figure}[htbp]
\begin{center}
\subfigure[][]{\includegraphics[width=0.85\linewidth]{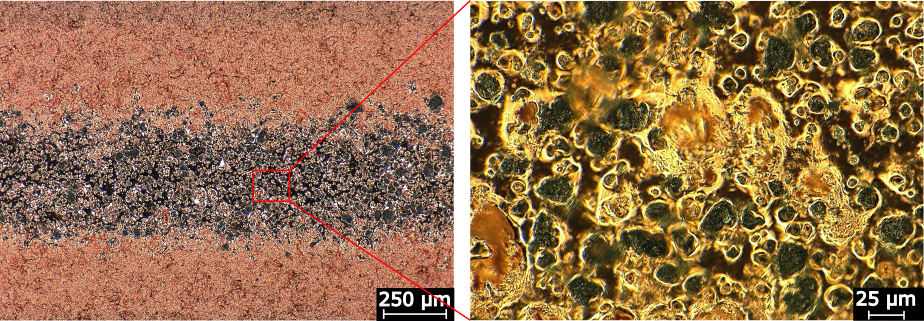}}
\subfigure[][]{\includegraphics[width=0.85\linewidth]{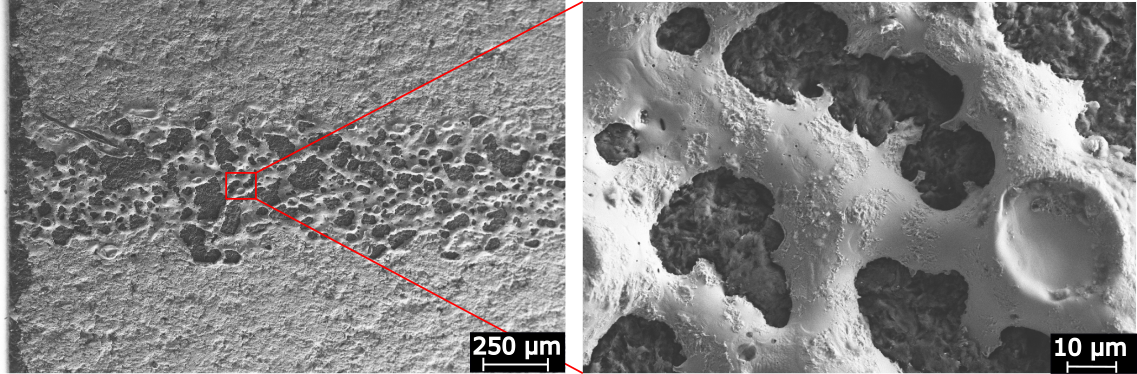}}
\end{center}
 \caption{Characteristic damage of the Cu coating by melting: a) optical microscope images of the Cu coated graphite jaw, acquired at 200x and 1000x, for the tilted impact at the highest apparent damage areas (area B in Fig.~\ref{fig16}) and b) SEM images at the entrance region for the tilted impact (area A).}\label{fig17}
\end{figure}

Images at medium and high magnification (x200 and x1000) in areas with significant apparent damage (Fig.~\ref{fig17}a) show a melted-droplet-like morphology of the coating with indications of solidification shrinkage. In addition, several SEM images (Fig.~\ref{fig17}b) were acquired to provide a higher contrast between the coating and substrate and show clearly uncoated areas, of $10$~$\mu m$–$100$~$\mu m$, with a kind of rounded pattern. Moreover, in some areas the coating exhibited some craters which could be the result of bursting bubbles or blisters.

\begin{figure}[htbp]
\begin{center}
\subfigure[][]{\includegraphics[width=0.55\linewidth]{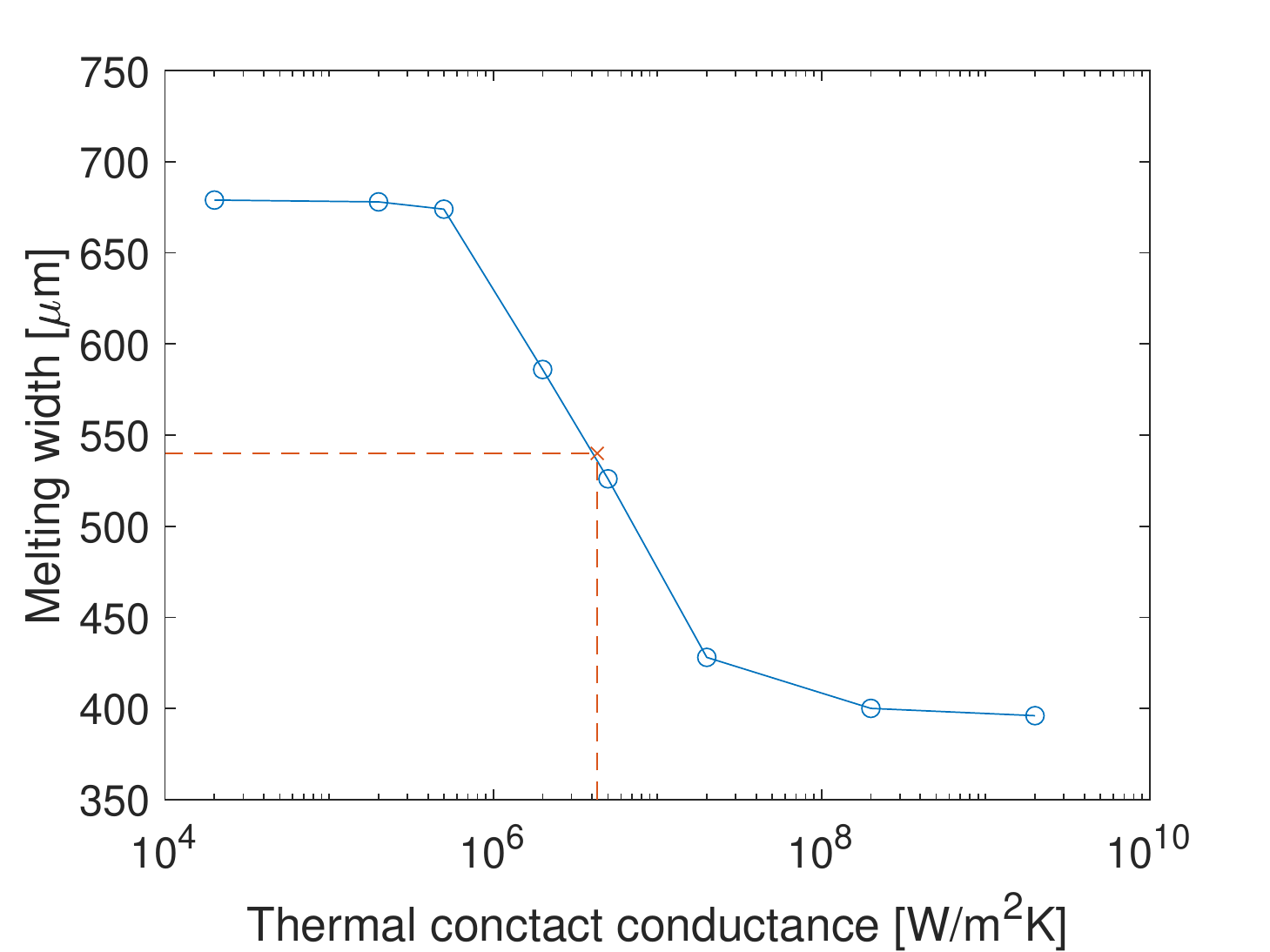}}
\subfigure[][]{
\begin{tabular}{p{0.64\textwidth} p{0.34\textwidth}}
  \vspace{0pt} \includegraphics[width=0.53\textwidth]{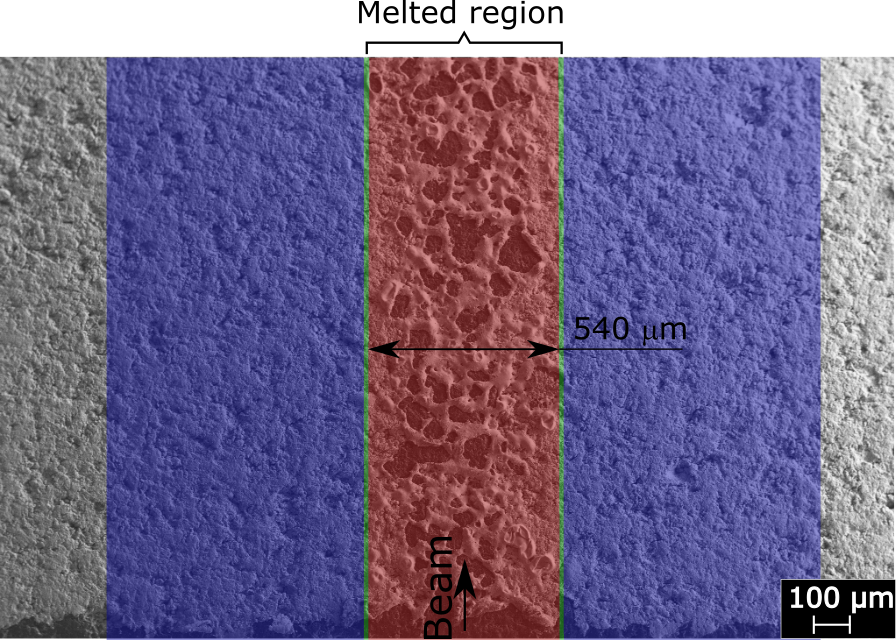} &
  \vspace{0pt} \includegraphics[width=0.25\textwidth]{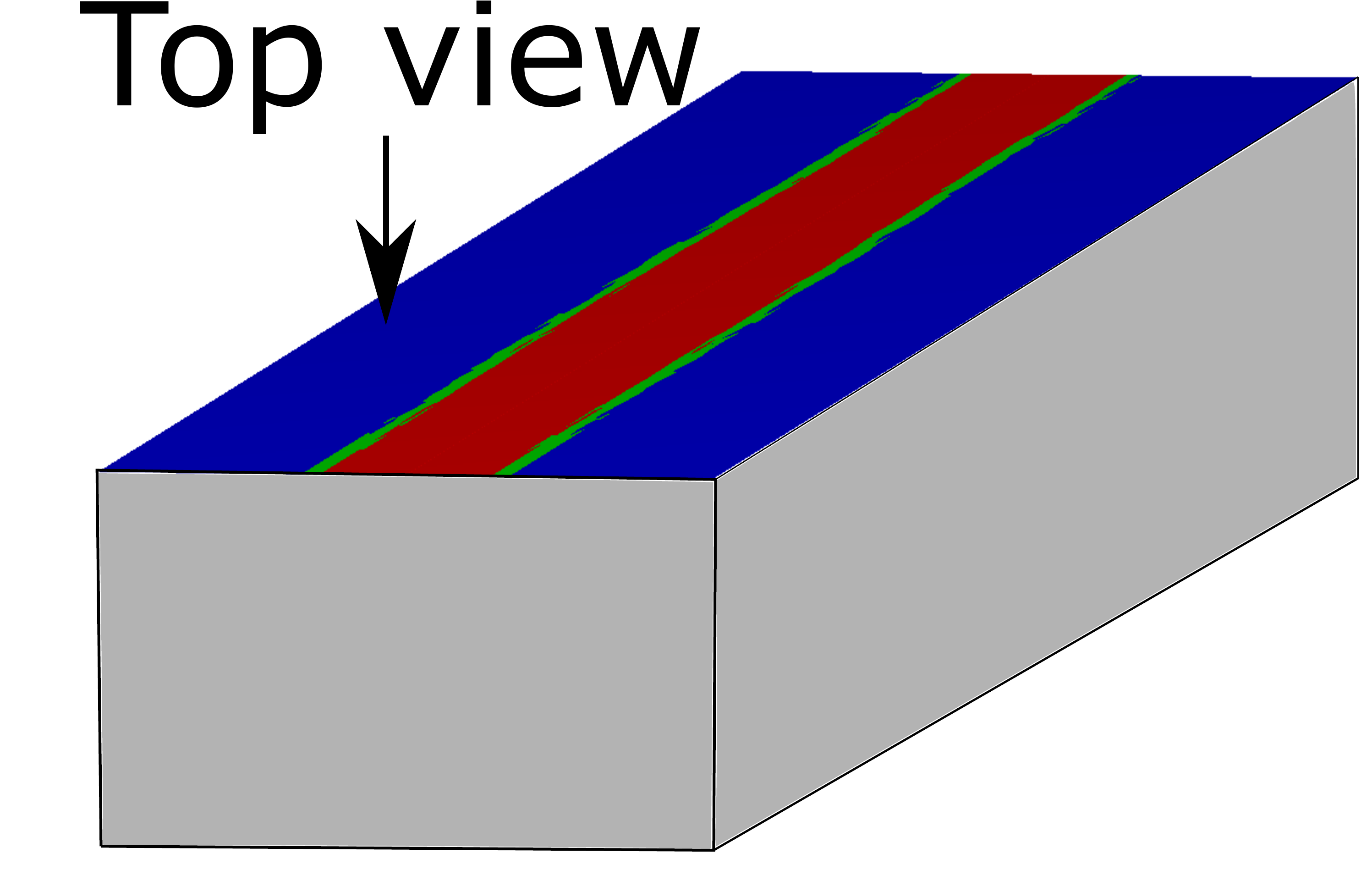}
\end{tabular}
}

\end{center}
 \caption{a) FEA findings for melted width of the copper coating at the beam entrance of the Cu-coated graphite block assuming a grazing impact as a function of the TCC (comparison of numerical and experimental findings at the orange marker). b) Superposition of the SEM image (black and white image) and numerical simulation (colour image, where melted region is in red and non-melted region is in blue) of the Cu coating (top view) for a $TCC_{Cu-Gr}^*=4.30\cdot 10^{6} [W/m^2K]$ (orange marker in previous graph a), showing the good agreement between simulations and experimental observations in terms of melted width.}\label{fig18}
\end{figure}

Numerical studies, presented in Sec.~\ref{Thermal_Analysis_Coating}, predicted the melting of the copper coating at the entrance region for the three kinds of impacts. As shown in Fig.~\ref{fig18}a, the size of the melting region is related to the TCC between coating a substrate, such that the width of the melting region diminishes with the TCC. The two asymptotic extremities of the curve, at low and high TCC values, correspond to a quasi-insulated and perfectly bonded contact, respectively.  A good agreement of the melting width between numerical and experimental observations can be found for a value of $TCC_{Cu-Gr}^*=4.30\cdot 10^{6}$~$[W/m^2K]$ (Fig.~\ref{fig18}b). This value indicates a good thermal contact and confirms the satisfactory bonding between coating and substrate. 

\subsubsection{Molybdenum coating}

\begin{figure}[htbp]
\begin{center}
\includegraphics[width=0.9\linewidth]{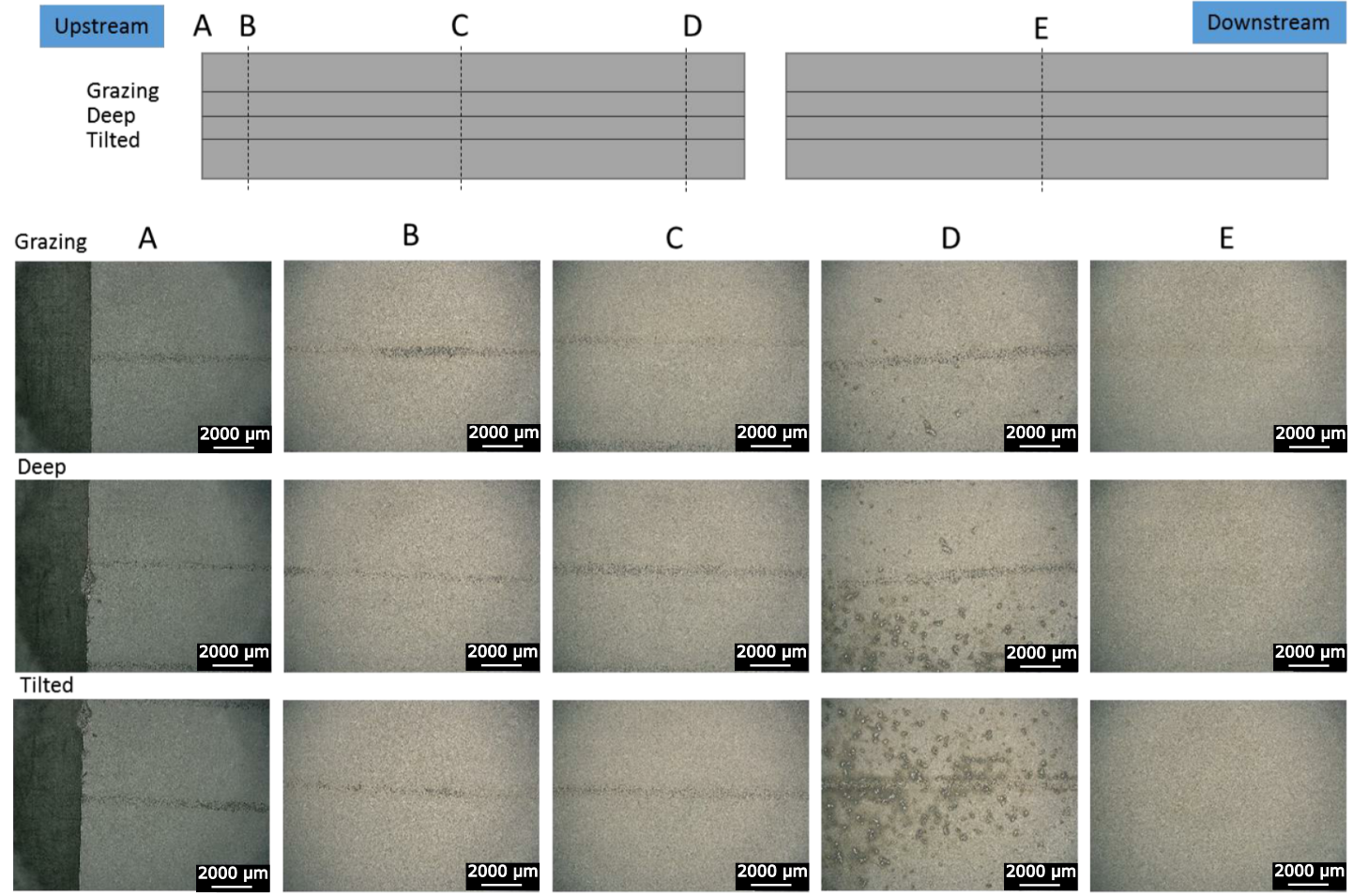}
\end{center}
 \caption{Optical microscope images at low resolution of the Mo-coated graphite target acquired at different points along the beam impact axes, showing the overall status of the coating after irradiation. Damage of the coating is more pronounced in the first block. Refer to the schematic illustration for the location of the corresponding image.}\label{fig19}
\end{figure}

Fig.~\ref{fig19}) shows an overview of the status of molybdenum-coated graphite jaw after irradiation. The three beam impact lines are visible on the coating and are more pronounced on the upstream half of the jaw (i.e the first block). In general terms, the coating exhibits similar levels of damage for the three types of beam impacts. The position of the greatest damage however, is different: while for the grazing impact the greatest damage is in the first quarter of the upstream block (area A and B), for the deep and tilted impacts the greatest damage is shifted to the second and third quarter (area C). It should be noted that in area D, for tilted and deep impacts , the debris coming from the upper jaws is  responsible for this spot-like pattern, rather than the direct beam-induced phenomenon. 

\begin{figure}[htbp]
\begin{center}
\includegraphics[width=0.8\linewidth]{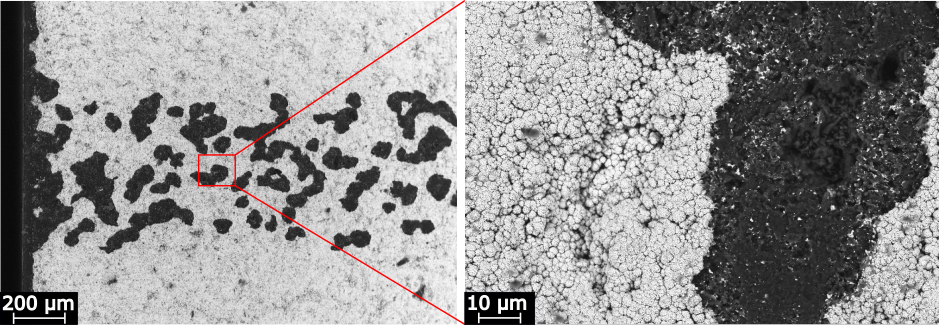}
\end{center}
 \caption{SEM images of the Mo-coated graphite jaw for a grazing impact at the entrance region (Fig.~\ref{fig19} area A). The dark areas indicate absence of coating due to the detachment phenomenon.}\label{fig20}
\end{figure}

SEM micrographs acquired at high magnification (Fig.~\ref{fig20}) reveal a  heterogeneous spot-like damage, with a characteristic size of about $50$~$\mu m$-$100$~$\mu m$, in which the coating detaches, most likely due to spallation. In these areas the entire thickness of the coating is removed. Unlike the copper coating, no signs of melting is found at any impact location. Based on this finding and recalling the previous thermal simulation performed on the molybdenum coated graphite (see Sec.~\ref{Thermal_Analysis_Coating}), a minimum threshold value of  $TCC_{Mo-Gr}^*=2.00\cdot 10^6 [W/m^2K]$ can be deduced, such that values higher than $TCC_{Mo-Gr}^*$ are compatible with these observations.   

\begin{figure}[htbp]
\begin{center}
\includegraphics[width=0.99\linewidth]{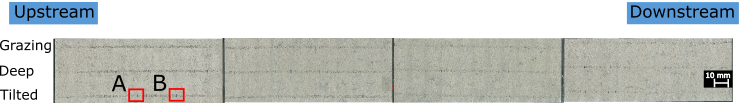}
\includegraphics[width=0.45\linewidth]{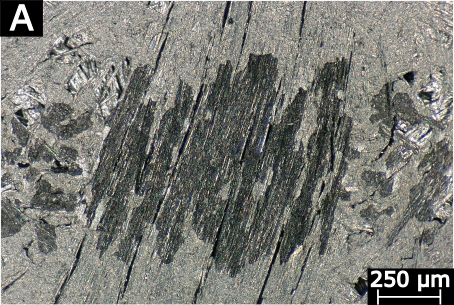}
\includegraphics[width=0.99\linewidth]{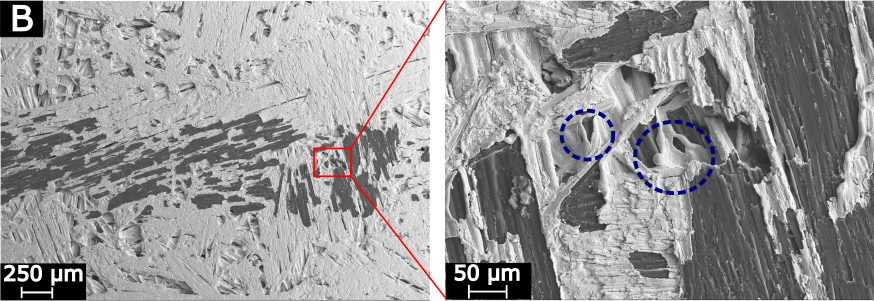}
\end{center}
 \caption{Characteristic damage of the Mo coated CfC for a deep impact on the first block: optical (A) and SEM (B) images at medium and high resolution. The images show the coating detachment following the fibre pattern of the substrate and the melting phenomena in localised areas (blue circles in B).}\label{fig21}
\end{figure}

In the case of the molybdenum-coated CfC, the coating damage is intermittent all along the jaw and is similar for the three types of beam impacts. The damage seems to be correlated with the substrate's fibrous structure. High-resolution micrograph images (Fig.~\ref{fig21}) show that the coating damage corresponds to a peel-off following the fibre pattern, this being more pronounced for the most exposed external layers or fibres. Moreover, minor signs of melting of the coating deposited around some isolated fibres and in interstitial spaces between fibres are visible, probably due to  low local thermal contact conductivity in that area.

\subsection{Analysis of the substrates}
\label{Substrate_analysis}

The structural integrity of the substrates (CfC, graphite and SiC-SiC) was analysed using X-ray computed tomography (Zeiss METROTOM 1500 CT scanner) that allows visualization of possible internal  damage with a voxel size in the range of $22$-$31$~$\mu m$ (depending on the material). The inspection was restricted to the region where the FLUKA simulations predicted the highest energy deposition in the substrate.

\begin{figure}[htbp]
\begin{center}
\subfigure[][]{\includegraphics[width=0.59\linewidth]{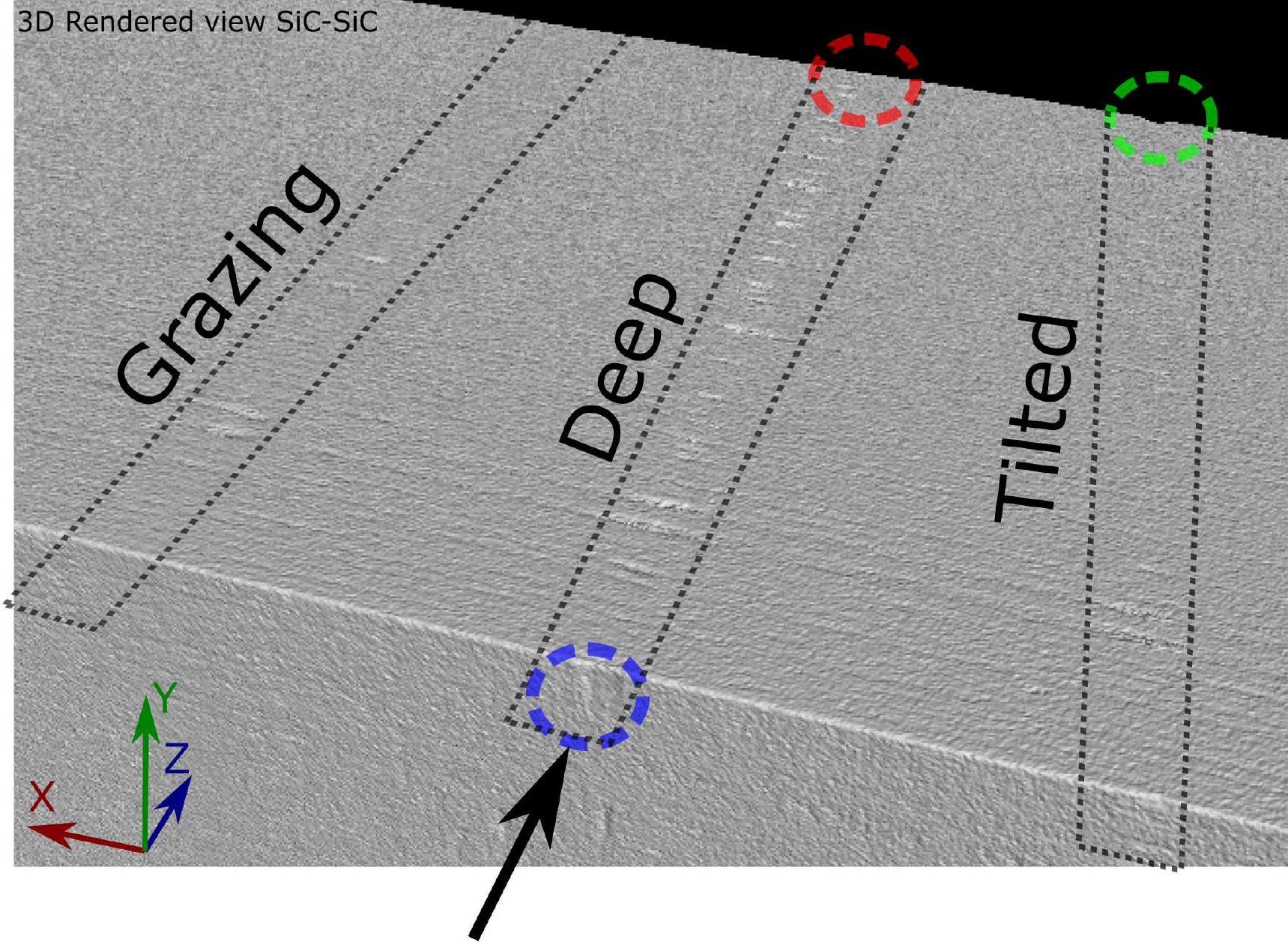}}
\subfigure[][]{\includegraphics[width=0.39\linewidth]{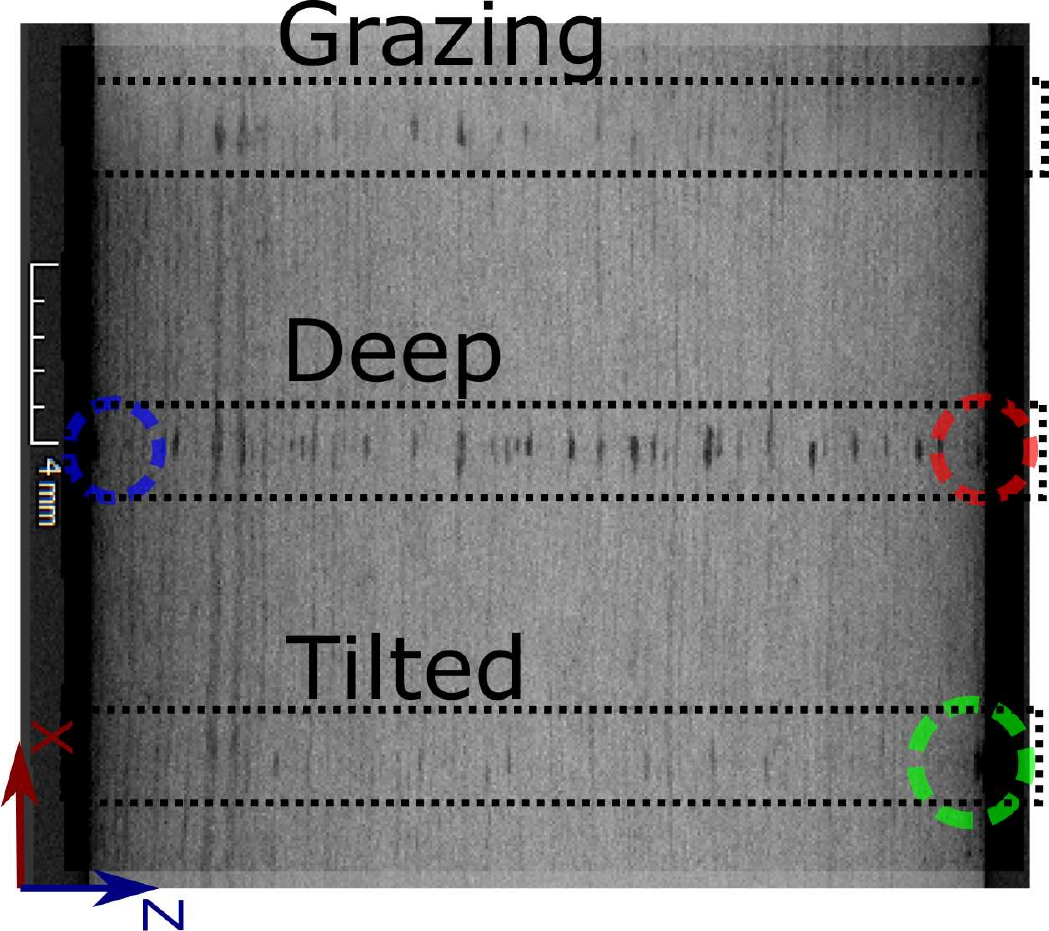}}
\subfigure[][]{\includegraphics[width=0.85\linewidth]{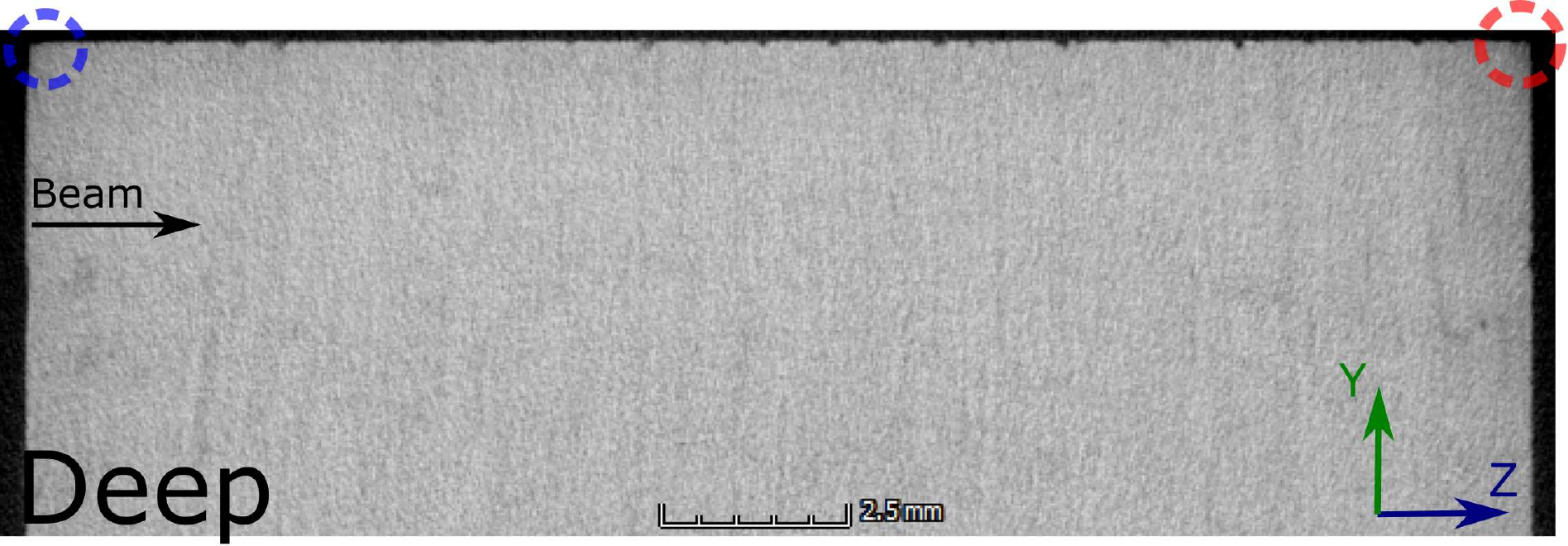}}
\end{center} 
 \caption{Computer tomography of the SiC-SiC block: a) 3D view, b) 2D top section view at a depth of 30 $\mu$m with respect to the top free surface and c) 2D right section view at the region of the deep impact. Coloured circles indicate the positions of craters at the entrance/exit faces.}\label{fig22}
\end{figure}

No internal cracks or damage were detected in the graphite and CfC substrates as a consequence of irradiation. Nevertheless, some damage can be observed at the free surface level of the SiC-SiC block, this being more pronounced for the deep beam impact (see Fig.~\ref{fig22}). The superficial damage decreases progressively with depth into the substrate, with no apparent signs of damage at $100$~$\mu$~m. Local craters at the entrance and exit faces of the block are also found in the deep and tilted beam impact positions on the block with a thickness of $22-50$~$\mu m$.  

\begin{figure}[htbp]
\begin{center}
\includegraphics[width=0.8\linewidth]{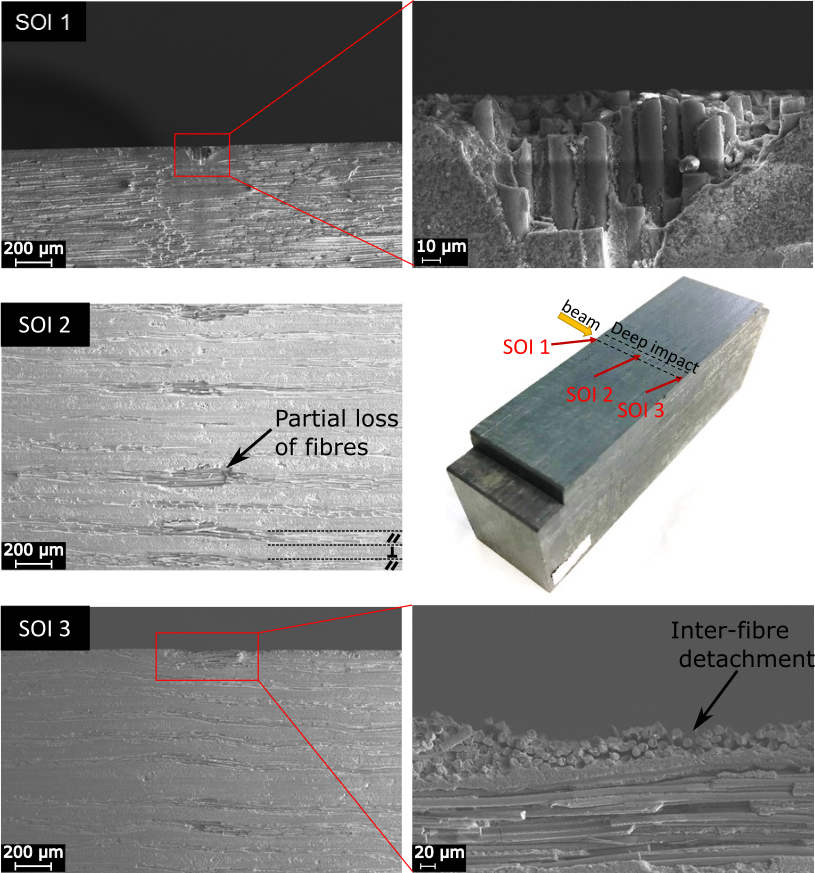}
\end{center}
 \caption{SEM micrographs of the SiC-SiC block for deep impact at three different sites of interest (SOI): beam entrance face (SOI~1); mid top view (SOI~2); and exit top view (SOI~3), showing the failure of the block. Symbols $\parallel$ and $\perp$ refers to fibres parallel and impinging to the top free surface, respectively.}\label{fig23}
\end{figure}

In order to extract more details about induced damage on surface morphology of the SiC-SiC block, SEM micrographs were acquired at the entrance, top free surface and exit, where the beam passed through (see Fig.~\ref{fig23}). Observing the entrance faces (SOI~1), the matrix is detached from the first layer of fibres, leaving them visible where the beam impacted. Additionally, some fibres of this layer are broken and there are signs of some interfibrillary debonding. This damage is observed for the deep and tilted beam impact zones, it is less intense in the tilted beam case and no apparent damage is observed in the grazing beam zone.

Regarding the top view (SOI~2), the damage is different depending on the orientation of the fibres. In the layers with fibres parallel to the free surface, fibre detachments are observed, with several fibres missing. In the layers with fibres impinging the free surface, damage is only observed in the proximity of the edge, with some inter-fibre detachment (SOI~3).

Observations based on examination of the targets are in qualitative agreement with the numerical simulation findings, where damage was expected for the SiC-SiC block and more intense damage in the case of deep impact. Nevertheless, it should be noted that the damaged region of the SiC-SiC is considerably smaller than the numerical predictions. As mentioned on the literature, SiC-SiC exhibits a quasi-ductile behaviour. This means that above at certain stress level, called  proportional limits~\cite{yb2012}, damage starts to appear inside the material, changing its behaviour (stiffness), but it is still able to withstand considerably higher stresses up to the fracture strength. This feature is not considered in the Tsai-Wu failure criterion used in previous numerical models (see Sec.~\ref{FEA_substrate}), where the failure is assumed to happen at the proportional limit (as a conservative assumption). Progressive degradation models should be then considered for predicting accurately the non-linear post damage evolution until the final fracture~\cite{garnich2009,zhang2021}. Moreover, although the Tsai-Wu criterion considers the typical orthotropic strength of composites, it cannot distinguish between the different types of composite failures, as observed in this experiment. Other criteria as proposed by Puck~\cite{matthias2012} or Hashin~\cite{hashin1980}, formulated at the level of the constituents, overcome this limitation. These models rely on complex characterizations, not currently available for this material, but could provide better damage estimation, which would be helpful for future studies.

\subsection{Impedance measurements}

Impedance measurements were performed on the three different types of coated blocks in order to evaluate quantitatively the implications of the observed coating damage on the in-plane resistivity. Several techniques have been proposed in the literature - such as the Four-point method, Eddy Current Testing and  Resonant Cavity Method~\cite{heaney2003,198942,juillard1990}. The Resonant Cavity method has been found to be accurate and especially suitable for thin coating films on a substrate and was therefore employed in the present work~\cite{coatings10040361}.

The Resonant Cavity Method uses an open copper cavity that has been designed to work in a specific resonant mode (typically the $H_{011}$), selected for enhancing the accuracy of the method. The cavity is connected to a Vector Network Analyzer (VNA) calibrated at the operational frequency. The sample to be analyzed is placed on the open side, working as an end cap and closing the cavity. As a result, the sample provokes a change in the wall resistivity and therefore a change in the Quality factor of the cavity (Q-factor). The resistivity of the sample is measured indirectly by measuring the relative change of the Q-factor with respect to a given reference material~\cite{coatings10040361}.% (copper end cap), which can be correlated by the fitting curve $Q/Q_{ref}=(a+b)/(a\sqrt(\rho/\rho_{ref})+b)$, where $a$ and $b$ are the energy deposited in the end cap and the rest of the cavity, and $Q_{ref}$ and $\rho_{ref}$ are the Q-factor and resistivity of a reference material, respectively~\cite{coatings10040361}.

\begin{figure}[htbp]
\begin{center}
\includegraphics[width=0.99\linewidth]{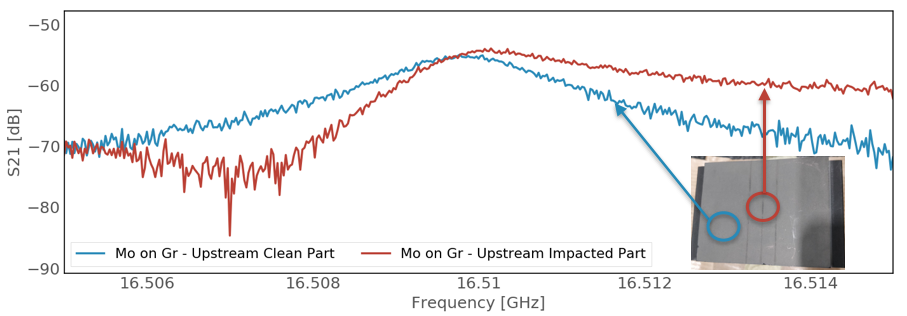}
\end{center}
 \caption{Measured transmission scattering parameters (S21) versus frequency for the Mo-coated graphite target at the entrance region, on the impacted (red curve) and non-impacted (blue curve) areas, showing an evident distortion of the S21 curve when measuring the damaged surface. }\label{fig24}
\end{figure} 

Prior to the measurements, the equipment and test method was calibrated with several metals. In addition the cavity was cleaned with ethanol to avoid surface contamination. For comparison purposes the measurements were made on the impacted and the non-impacted areas of the coatings. Figure~\ref{fig24} shows the measured transmission scattering parameter (S21) versus frequency for the Mo-coated graphite target, from which the Q-factor and resistivity could be deduced. Similar observations hold for copper-coated graphite blocks. The measurement results show an evident distortion of the S21 curve on the impacted area that could be related to the perturbation of a lower frequency mode sensitive on the degraded surface contact on the measured area (red circle in Fig.~\ref{fig24}). Due to the distortion of the transmission scattering parameter, it is not straightforward to quantify the resistivity of the damaged surface, whereas about 800~n$\Omega$m are estimated for the non-impacted Mo area~\cite{AdnanWP2}. Given the local nature of the beam impact region, this is retained not to represent an issue from the impedance point of view: orbit bumps and/or BIDs' jaw position adjustments can be done in the eventuality of an impact on the absorbing materials.
No results could be extracted for the Mo-coated CfC block, being the employed technique not suitable to the rough surface inherent to the fibrous structure of this material.

\section{Conclusions}
\label{Conclusion}
In the present work, the performance of coated-absorbing blocks exposed to LHC - like proton beam impacts, representative of the LHC TDIs' operating conditions, was assessed using a combination of practical beam test experiments and numerical analysis techniques. Before exposure to the beams, isostatic graphite targets showed a good homogeneity of the coating deposition for both Cu and Mo coatings, with a thickness close to the technical specifications. Coating adhesion for these targets as well as for Mo-coated CfC was found to be satisfactory. 

Post irradiation examinations revealed that graphite and CfC substrates displayed no internal damage as a result of any of the types of beam impact. SiC-SiC blocks displayed superficial damage for all impacts and had craters at the entrance and exit faces for deep and grazing impacts. In general terms, damage of the SiC-SiC was linked to matrix detachment, breakage of the fibres parallel to the top surfaces and debonding between matrix and fibres. These observations are coherent with the numerical analysis findings, although this analysis predicts larger damage most probably due to conservative assumptions.

Optical microscopy and SEM observations of the coatings show clear damage to both molybdenum and copper coatings. The Cu coating was melted around the beam impact region, more noticeably for the grazing and tilted impacts, as predicted by the numerical models. No sign of melting was found for the Mo coating on the graphite blocks, although it was found spot-like detachments,  most probably due to spallation phenomena. On the Mo-coated CfC blocks, a similar detachment was seen following the fibre pattern and some minor signs of melting were also observed in between fibres. 

Impedance measurements on both damaged and undamaged surfaces of coated graphite blocks were performed showing a similar distortion on the transmission scattering parameter measured on the impacted area of Mo and copper coated surfaces. In both cases, the overall performance in case of an eventual beam impact can be restored with appropriate orbit bumps or jaw position adjustments. 

\begin{acknowledgments}
The authors would like to acknowledge the support of CERN's Sources, Targets and Interactions (STI) Group, the Accelerator Consolidation Project at CERN, as well as M. Wendt and F. Caspers. This program is partially supported by JSPS KAKENHI Grant Number JP16H03994. This support is gratefully acknowledged. The research leading to these results has received funding from the transnational access activity ARIES which is co-funded by the European Union's Horizon 2020 Research and innovation programme under Grant Agreement no. 730871.
\end{acknowledgments}
\bibliography{bibfile}
\bibliographystyle{unsrturl}

\end{document}